\documentclass[18pt,a4paper,onecolumn,fleqn,usenatbib,useAMS,referee]{mnras}

\usepackage{graphicx}	
\usepackage{amsmath}	
\usepackage{amssymb}	
\usepackage{multicol}   
\usepackage{bm}		    
\usepackage{pdflscape}	
\usepackage{float}
\usepackage{caption}

\usepackage{placeins}

\usepackage[T1]{fontenc}
\usepackage{ae,aecompl}

\usepackage{newtxtext,newtxmath}

\title[Thermal Conductivity]{Thermal Conductivity measurements of macroscopic frozen Salt Ice analogs of Jovian Icy moons in support of the planned JUICE mission}

\author[Gonz\'{a}lez D\'{i}az et al.]{
C. Gonz\'{a}lez D\'{i}az$^{1}$,\thanks{E-mail: 
\href{mailto:cgonzalez@cab.inta-csic.es}{cgonzalez@cab.inta-csic.es}; 
\href{maito:munozcg@cab.inta-csic.es}{munozcg@cab.inta-csic.es}}
S. Aparicio Secanellas$^{2}$,
G. M. Mu\~noz Caro$^{1}$\footnotemark[1], \newauthor
J.J. Anaya Velayos$^{2}$, 
H. Carrascosa$^{1}$, 
M. G. Hern\'{a}ndez$^{2}$, 
V. Muñoz-Iglesias$^{1}$,\newauthor
\'{A}. Marcos-Fern\'{a}ndez$^{3}$, 
O. Prieto-Ballesteros$^{1}$, 
R. Lorente$^{5}$,\newauthor
O. Witasse$^{6}$ and
N. Altobelli$^{5}$ \\
$^{1}$Centro de Astrobiolog\'{\i}a (CSIC-INTA), Ctra. de Ajalvir, km 4, Torrej\'on de Ardoz, 28850 Madrid, Spain.\\
$^{2}$Instituto de Tecnolog\'{\i}as F\'{\i}sicas y de la Informaci\'on, Leonardo Torres Quevedo, (ITEFI-CSIC), c/ Serrano 144, 28006 Madrid, Spain.\\
$^{3}$Instituto de Ciencia y Tecnolog\'{\i}as de Pol\'{\i}meros
(ICTP-CSIC), Juan de la Cierva, 3, 28006 Madrid.\\
$^{4}$European Space Agency (ESA) - ESAC Camino Bajo del Castillo s/n Villafranca del Castillo, 28692, Villanueva de la Cañada (Madrid), Spain.\\
$^{5}$European Space Agency (ESA), European Space Research and Technology Centre (ESTEC), Noordwijk, Netherlands}

\date{Accepted XXX. Received YYY; in original form ZZZ}

\pubyear{2020}

\begin{document}
\label{firstpage}
\pagerange{\pageref{firstpage}--\pageref{lastpage}}
\maketitle

\begin{abstract}\label{abstract}
The study of thermal properties of frozen salt solutions representative of ice layers in Jovian moons is crucial to support the JUpiter ICy moons Explorer (JUICE) (ESA) and Europa Clipper (NASA) missions, which will be launched in the upcoming years to make detailed observations of the giant gaseous planet Jupiter and three of its largest moons (Ganymede, Europa, and Callisto), due to the scarcity of experimental measurements. 
Therefore, we have conducted a set of experiments to measure and study the thermal conductivity of macroscopic frozen salt solutions of particular interest in these regions, including sodium chloride (NaCl), magnesium sulphate (MgSO$_4$), sodium sulphate (Na$_2$SO$_4$), and magnesium chloride (MgCl$_2$). Measurements were performed at atmospheric pressure and temperatures from 0 to -70$^{\circ}$C in a climatic chamber. Temperature and calorimetry were measured during the course of the experiments. An interesting side effect of these measurements is that they served to spot phase changes in the frozen salt solutions, even for very low salt concentrations. A small sample of the liquid salt-water solution was set aside for the calorimetry measurements.
These experiments and the measurements of thermal conductivity and calorimetry will be valuable to constrain the chemical composition, physical state, and temperature of the icy crusts of Ganymede, Europa, and Callisto.
\end{abstract}

\section{Introduction} \label{Introduction}
Measurements of thermal conductivity in frozen salt solutions as analogs of Jovian icy moons are very scarce in the literature. These moons will be visited in the coming years by the JUICE (ESA) and Europa Clipper (NASA) missions. JUICE is the first large-class mission in ESA's Cosmic Vision 2015-2025 programme. Planned for launch in 2022 and arrival at Jupiter in 2029, it will spend at least three years making detailed observations of Jupiter and its environment.

In this work, we propose to study the thermal conductivity and calorimetry of frozen solutions containing one of these salts: sodium chloride (NaCl), magnesium sulphate (MgSO$_4$), sodium sulphate (Na$_2$SO$_4$), and magnesium chloride (MgCl$_2$). These samples were selected as ice analogs in Jovian moons.

The focus of JUICE is to characterise the conditions that may have led to the emergence of habitable environments among the Jovian icy satellites, with special emphasis on the three ocean-bearing worlds: Ganymede, Europa, and Callisto \citep{KIVELSON_2000}.

Ganymede is the primary scientific target of the mission. With a diameter of about 5260 km, it orbits Jupiter at an average distance of 1.070.400 km. Researchers believe that tidal heating likely occurred on Ganymede and drove tectonic activity \citep{CAMERON_2018}. There is evidence for a subsurface layer of salt water under its icy surface \citep{KIVELSON_2000,COLLINS1998,PAPPALARDO1998,Golombek1981,SHOWMAN2004,Prockter2000,PATTERSON2010,Parmentier1982Natur,McCord2001,Noll1996,Hendrix1999,Clark1980,HEGGY2017,McCord2001,McCord1997,McCord1998JGR,Pettinelli2015,Sohl2010}. The depth limits of these oceans are unknown and how it interacts with both the deep interior of Ganymede and the icy crust above it. Finding out more about Ganymede's ice and liquid layers, including its composition and conductivity, is a main objective of JUICE since the ocean might be habitable. The mission will use an ice-penetrating radar to probe the moon's subsurface structure down to a depth of about 9 km.
Europa is the second target of JUICE. With a diameter of 3121.6 km and average distance from Jupiter of 670.900 km, tidal heating is important in this moon. It hosts an ocean beneath its frozen outer shell and offers the best habitability conditions within the Jovian satellites \citep{Carlson1996,Carlson1999a,Carlson1999,DALTON2013,Fink1973,DALTON2005,Fink1973,Hansen2008,KARGEL1991,Kargel2000,Marion2005,McCord1998,ORLANDO2005,PRIETOBALLESTEROS2005}. JUICE will explore the chemistry of Europa's water-vapour potential plumes emerging from its young ice crust \citep{Sparks_2016,Roth_2014}. Active zones in Ganymede and Europa could provide a way to exchange material and heat between the moon's surface and the ocean beneath \citep{GRASSET_2013}.
JUICE will also study Callisto. This moon has a diameter of 4821 km and average orbital radius of 1.883.000 km. Callisto experienced lower tidal heating than Ganymede and is exposed to lower radiation field than the other Galilean moons. The presence of an ocean beneath its rocky and ice surface is not as clear \citep{Greeley2000P&SS,ZIMMER2000,Schenk1993JGR,Schenk1995,Schenk2002Nature,Showman1999,Clark1980,HEGGY2017,McCord1997,McCord1998JGR,Pettinelli2015,Sohl2010}. Callisto has an old cratered surface and is geologically stable compared to the other Galilean moons \citep{ANDERSON_20011,KUSKOV_2005}. For more details on JUICE mission, see \href{https://sci.esa.int/web/juice}{https://sci.esa.int/web/juice}.
The JUICE and Europa Clipper missions incorporate a radar to explore the surface and subsurface of the moons. Clipper will also measure temperature changes on Europa's surface and a mass spectrometer to determine the surface composition. Further information can be found in \href{https://www.jpl.nasa.gov/missions/europa-Clipper}{https://www.jpl.nasa.gov/missions/europa-Clipper}. 
In this paper, the measurements of temperature and thermal conductivity of ice analogs containing salts with the different concentrations will serve as input for thermodynamical models of Jovian icy moons. These models will be developed to interpret the volume of data from these missions.
The applicability of the data reported in this paper to Jovian icy moons requires an extrapolation of the curves to lower temperatures. Unfortunately, there is no literature reporting similar measurements at lower temperatures that we could use as a reference. However for the pure water case there are some works that measure the thermal conductivity at temperatures below -50$^{\circ}$C \citep{KLINGER80, Slack_1980, Andersson_2005}. These authors report an exponential increase of thermal conductivity as the temperature decreases.
In addition to the temperature thickness of the different liquid/ice layers of the moon, the thermal conductivity is required as input for estimation of heat transfer as a function of depth. The thermal conductivity value depends on the composition and physical properties of the salt ice and changes as a function of the temperature.  
\section{Experimental protocol}
\label{Experimental protocol}
In this section the sample production is described and the characterization of the macroscopic ice samples containing salts. The protocols for temperature, thermal conductivity and calorimetry measurements are presented.
Macroscopic frozen salt solutions are of special concern in three of the largest Jovian Moons with an ice crust: Ganymede, Europa and Callisto. Analog solutions were made according to the literature input requirements. The chemical components of the samples are: sodium-chloride (NaCl), magnesium-sulphate (MgSO$_4$), sodium sulphate (Na$_2$SO$_4$), and magnesium chloride (MgCl$_2$) with different concentrations, see Table~\ref{saltsystem}. The water used to fabricate the ice samples is “ultrapure” water of “Type 1” (Milli-Q).
The chemicals Sodium chloride 99.5-100.5\%, AnalaR NORMAPUR® ACS, Reag. Ph. Eur. analytical reagent, 
Magnesium chloride, anhydrous $\geq 98\%$, high purity, Sodium sulphate, anhydrous 98.5-101.0\%, AnalaR NORMAPUR® Reag. Ph. Eur. analytical reagent, and Magnesium sulphate, anhydrous 99-100.5\% USP salts were purchased from AVANTOR.
For each ice sample, temperature and thermal conductivity were measured. For the temperature measurement, 8 thermocouples of "type T", "class 1" were placed in 2 rods. The temperature was recorded every 30 seconds. For each rod, one was kept outside the sample to measure the temperature in the interior of the chamber and the other 3 were embedded in the sample at different depths (at 0, 5 and 10 cm from the ice surface level). The thermal conductivity was measured with a commercial device TEMPOS--thermal properties analyser from "METER Group". The TR-3 sensor of TEMPOS was used to measure the thermal conductivity $K$ of the frozen salt solutions. This single-needle TR-3 sensor measures thermal conductivity and thermal resistivity. The TR-3 is primarily designed for soil and other granular or porous materials. This single-needle sensor is based on the hot-wire probe method, consisting of a needle with a heater and temperature sensor inside. The sensor can measure temperatures in the range of  -50$^{\circ}$C to  +150$^{\circ}$C with a precision of 0.001$^{\circ}$C. The sensor dimensions are 100 mm long and 2,4 mm diameter. The thermal conductivity measurement range of this device is 0.1--4.0 $\pm$ 10\% W/(m·K). The peaks in the thermal conductivity measured during phase changes correspond to unrealistically high values of the thermal conductivity, for instance see Fig.~\ref{TemperatureH2O-1} between 153 and 162 hours, at 0$^{\circ}$C. These peaks served to spot phase changes as they are caused by a change in the temperature during the phase change that the sensor interprets as a large variation of the thermal conductivity. In addition, for solutions at the phase change temperature, convection induces temperature variations and the thermal conductivity measurements may have large uncertainties.
In Fig.~\ref{TemperatureH2O-1} to clarify the precision of the thermal conductivity sensor ($\pm$ 10\% W/(m·K)) relative to the overall measurements, three error bars were placed at the beginning, middle and near the end of the experiment. Since the error is always 10\% of the thermal conductivity measurement, the error bars for all the points have not been plotted to avoid data overload in the figure.
The error bars for the thermal conductivity temperature sensor have not been added since the error is only $\pm$0.001$^{\circ}$C. Moreover, the temperature is also doubly measured with two arrays of 4 thermocouples each (two thermocouples per position) to further reduce the temperature error. The impurities in the samples which could introduce any error in the thermal conductivity measurements are insignificant since they are negligible compared to the amount of water or salt used to manufacture them. For instance, for the eutectic sample of NaCl (23.16 wt.\% NaCl + H$_2$O) 5 litres of “ultrapure” water (Milli-Q) and 1510.7 g of NaCl were used.
The error of the balance used is 0.01 g, a clearly negligible value for the sample amounts used in our experiments that corresponds to 0.0006\% of the salt weight, errors of this order are estimated for the other experiments.

\begin{table}
	\begin{center}
    \begin{tabular}{||c|c|c||}
    	\hline
    	\hline
    	\textbf{Salt System }& \textbf{Eutectic temperature ($^{\circ}$C) }& \textbf{Initial concentration (wt.\%) }\\
    	\hline
    	\hline
    	NaCl & -21.1 & 1.2, 23.16 (eutectic) in Fig.~\ref{phasediagramNaCl+H2O} \\
    	\hline
    	MgSO$_4$ & -3.7 & 17 (eutectic) in Fig.~\ref{phasediagramMgSO4H2O}\\
    	\hline
    	Na$_2$SO$_4$ & -1.3 and -3.5 & 4.15 (eutectic 1), 12.8 (eutectic 2), 17 in Fig.~\ref{phasediagramNa2SO4H2O}\\
    	\hline
    	MgCl$_2$ & -33.2 & 21 (eutectic)  in Fig.~\ref{phasediagramMgCl2+H2O}\\
    	\hline
    	\hline
    \end{tabular}
    \caption{Salt systems and concentrations selected for this study based on phase diagrams reported in the literature \citep{PhasediagramNaClandMgCl2, PhasediagramMgSO4, PhaseNA2SO4} and ref. therein.}
    \label{saltsystem}
    \end{center}
\end{table}
\FloatBarrier

\begin{figure}
	\centering
	\includegraphics[width=\hsize]{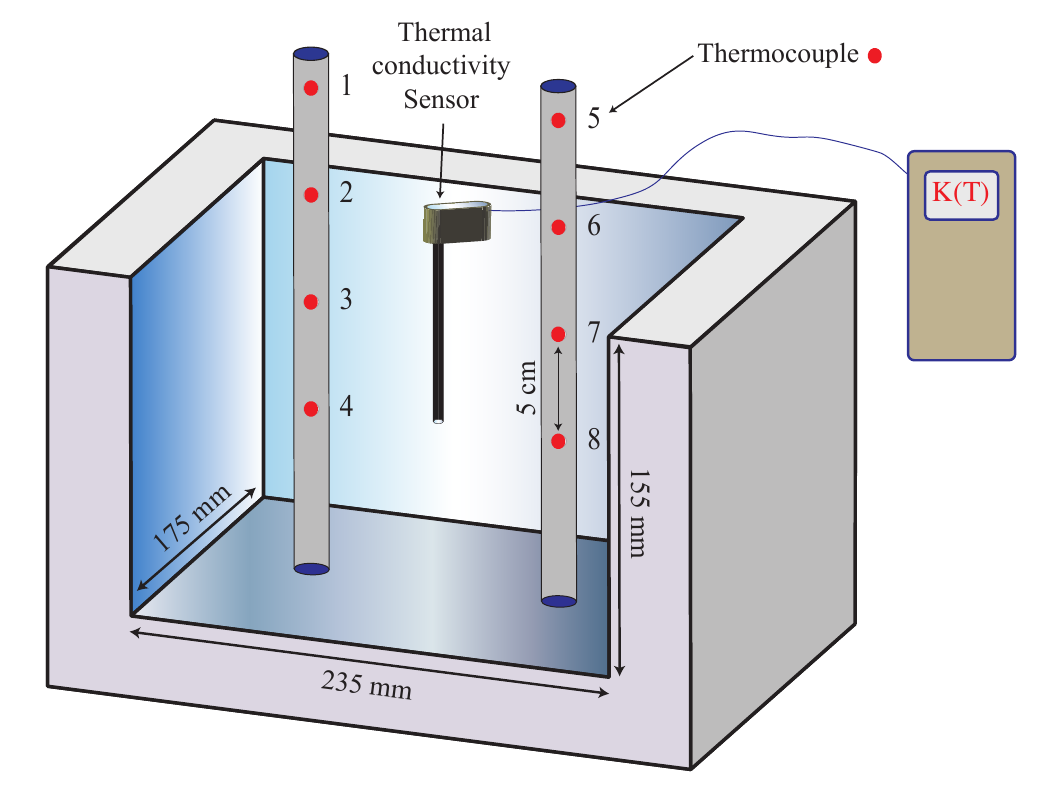}
	\caption{Dimensions of the interior of the EPS box and schematic view showing the positioning of the thermocouples and thermal conductivity sensor.}
	\label{AI_drawing}
\end{figure}

\begin{figure}
	\centering
	\includegraphics[width=\hsize]{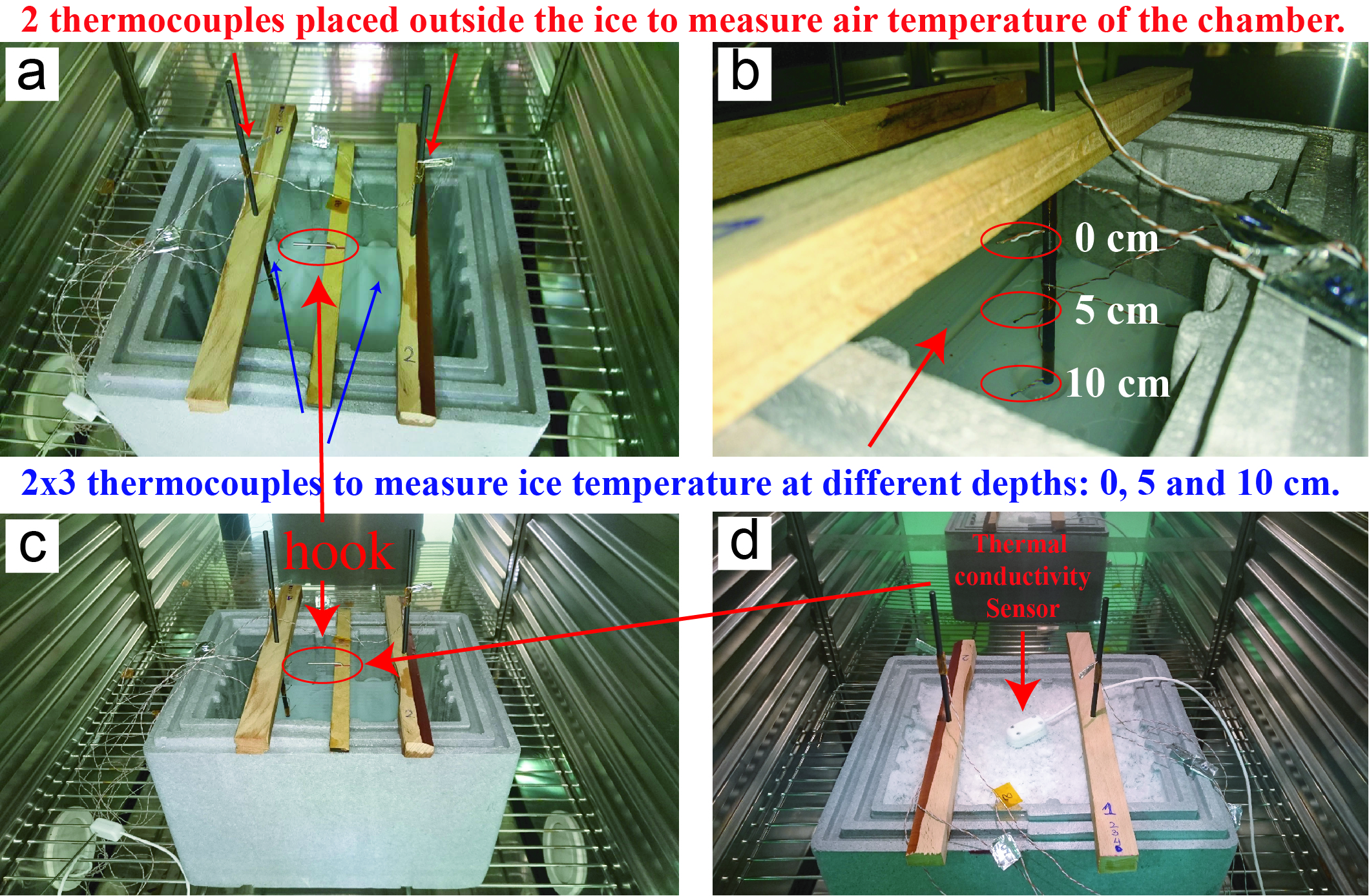}
	\caption{Schematic view of the ice sample formation using the climatic chamber and the EPS box. a) Salt solution in the EPS box placed inside the climatic chamber showing the thermocouple arrays, b) thermocouples positioned at different depths in the salt solution, c) the hook used as guide, to be replaced by the thermal conductivity sensor, and  d) thermal conductivity sensor in the salt solution.}
	\label{Schematic view of the ice sample formation using the climatic chamber and the EPS box}
\end{figure}

A solution containing a given salt concentration was prepared after stirring and placed in the interior of an expanded polystyrene (EPS) dry box, see Figs.~\ref{AI_drawing} and ~\ref{Schematic view of the ice sample formation using the climatic chamber and the EPS box}. Fig.~\ref{AI_drawing} shows the dimensions of the interior of the expanded polystyrene (EPS) dry box and schematic view showing the two rods with positioning of the eight thermocouples and the thermal conductivity sensor used to measure the temperature and the thermal conductivity of the ice samples.
Besides, the thermal conductivity and the temperature, a small sample of the liquid salt-water solution was set aside for the calorimetry measurements to determine the changes in energy of the samples by measuring the heat exchanged with the surroundings when the temperature is increasing/decreasing. Heat flow variations with the temperature were measured with the µDSC7 evo calorimeter (SETARAM Instrumentation), with a signal-to-noise ratio of 0.4 $\mu$W and a resolution of 0.02 $\mu$W. The reference cell was left empty in all the runs. The protocol for the calorimetry test is the following: 1) the sample solution is cooled down from 25$^{\circ}$C to -40$^{\circ}$C at the speed of -5$^{\circ}$C/min, then, it is kept at -40$^{\circ}$C during 20 minutes and finally, the sample solution is heated up from -40$^{\circ}$C to 20$^{\circ}$C at 1$^{\circ}$C/min.
For the MgCl$_2$, since the eutectic point at -33.2$^{\circ}$C, is closer to the limits of our previous calorimeter device, -40$^{\circ}$C, the thermal transitions were analyzed by Differential Scanning Calorimetry (DSC) on a Mettler Toledo DSC 822e calorimeter (Schwerzenbach) equipped with a liquid nitrogen accessory. A sample weighing approximately 20 mg in an aluminium pan was cooled from 25$^{\circ}$C to -90$^{\circ}$C at a rate of 10$^{\circ}$C/min, maintained for 7 min at this temperature and reheated from -90$^{\circ}$C to 25$^{\circ}$C at a rate of 10$^{\circ}$C/min. All scans were carried out under a constant nitrogen purge of 30 mL/min. Phase changes were taken as the onset of the transition for sharp peaks, and as the extremum peak temperature for broad bands, respectively \citep{LIN_2013}.

To perform the experiments a climatic chamber was used, manufactured by Votsch (model: VTR 7033/S). The chamber temperature can operate in the temperature range from -70$^{\circ}$C  to 120$^{\circ}$C and temperature ramps of 7.5 K/min (heating) and 6.5 K/min (cooling) can be conducted. The humidity ranged between 10\% and 98\% and the refrigerant R404A was used.  
The expected ice temperatures in Jovian moons are well below -100$^{\circ}$C. Due to experimental limitations, measurements were performed at atmospheric pressure and temperatures from 0 to - 70$^{\circ}$C, and the thermal conductivity could only be measured from 0$^{\circ}$C to -50$^{\circ}$C. The climatic chamber  does not allow inclusion of volatile species in the ice. Large blocks of salt ice can be grown inside the climatic chamber and the scientific equipment (thermocouples and the thermal conductivity sensor) is connected to the chamber via a port, see Fig.~\ref{Schematic view of the ice sample formation using the climatic chamber and the EPS box}. It allows introducing samples and sensors inside the chamber while the electronics remain out of the climatic chamber. 
Fig.~\ref{Schematic view of the ice sample formation using the climatic chamber and the EPS box} shows photographs of the ice sample formation using the climatic chamber and the EPS box. A hook with similar dimension as the thermal conductivity sensor was inserted in the solution as guide to the thermal conductivity sensor. Once the salt solution is frozen, this hook is replaced by the thermal conductivity sensor. 

As soon as the EPS box containing the salt solution is placed in the interior of a climatic chamber, the temperature of the climatic chamber is set to the minimum temperature range, around -70$^{\circ}$C during 1-2 days, depending of the salt solution, to permit freezing of the complete sample volume. This can be seen for instance in Fig.~\ref{TemperatureH2O-1}, where the thermocouple sensors 1 (blue line) and 5 (magenta line) are shown. These sensors are located outside the EPS box and were used to record the ambient temperature of the climatic chamber, displaying a plateau at around -70$^{\circ}$C during the first 24 hours of the experiment. When the whole salt solution was frozen, a heating ramp was programmed until the salt solution reached the eutectic temperature. 
To allow potential structural changes in the ice that may require more time, we introduced "steps" in temperature, where the temperature was kept intentionally constant at a given temperature value, for instance see Fig.~\ref{TemperatureH2O-1}. The duration in time of the "steps" was not the same for all the experiments or salt solutions studied. Not all the salt solutions took the same amount of time to reach the temperature due to their lower or higher thermal conductivity and specific heat.

\section{Experimental results} \label{Experimental results}
\subsection{Pure H$_2$O ice}
\label{pure water}
\begin{figure}
	\centering
	\includegraphics[width=\hsize]{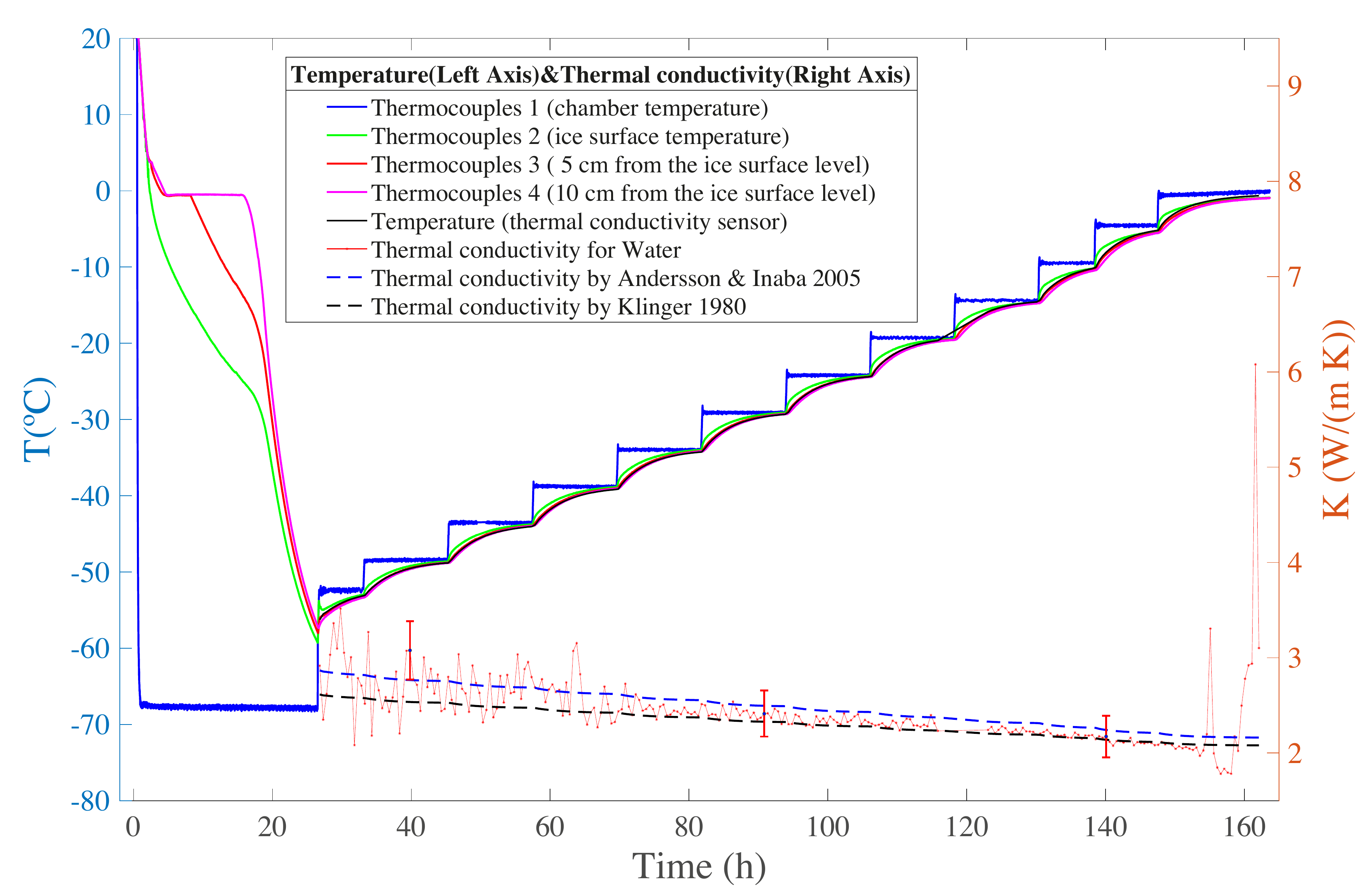}
	\caption[]{Thermal conductivity and temperature values measured by thermocouples and thermal conductivity meter for pure H$_2$O ice as a function of time. Previously published thermal conductivity values by \cite{Andersson_2005} (dashed blue trace) and by \cite{KLINGER80} (dashed black trace). To link the temperature values to their corresponding thermal conductivity values, for instance at 100 h, an imaginary vertical line should be projected from the thermal conductivity value to the temperature values on top.}
	\label{TemperatureH2O-1}
\end{figure}

This experiment served to test the performance of the setup as the results can be compared with the literature values. The phase change of water from liquid to solid can be seen in Fig.~\ref{TemperatureH2O-1}, where the thermocouple sensors show a plateau at 0$^{\circ}$C during the first hours of the experiment. The duration of the plateau increases as the thermocouples are placed deeper in the interior of the EPS box. This is the case of thermocouples 4 and 8 placed at a depth of 10 cm from the ice surface level. At these positions the water solution remains in the liquid state for about 16 hours. These plateaus or lower rate of cooling illustrate that the heat is being released as water ice continuously freezes, which is the result of slowing down the cooling rate of water. 

Once the water solution has fully solidified, heat is no longer released by solidification and the water ice cools down rapidly. Temperature drops from 0$^{\circ}$C to -60$^{\circ}$C in 10 hours.
Upon warming, the water ice melts. This is shown by a long period of constant temperature due to the endothermic melting. The temperature of this plateau is at 0$^{\circ}$C in the last step of the experiment between 156 and 165 hours.

Fig.~\ref{TemperatureH2O-1} also shows the measurements of the thermal conductivity of a pure water ice sample down to -50$^{\circ}$C to verify the correct functioning of the experimental setup. The temperature measured with the conductivity meter was quantified as "very similar" to the temperature measured with the deepest thermocouples. The thermal conductivity values are in good agreement with the literature values in this temperature range \citep{Slack_1980,KLINGER80, Andersson_2005, Carey2018, Bonales2017} as could be stated by the blue (thermal conductivity by \cite{Andersson_2005}) and black (thermal conductivity by \cite{KLINGER80}) dashed line in the Fig.~\ref{TemperatureH2O-1}.

Fig.~\ref{TemperatureH2O-1} shows few peaks occurring around 0$^{\circ}$C in the thermal conductivity measurement (red traces). These peaks coincide with the phase change from solid to liquid in the H$_2$O ice. 
The temperature ramp displays "steps" in Fig.~\ref{TemperatureH2O-1} because the temperature was kept intentionally constant at a given temperature value to allow potential structural changes in the ice that may require more time, see Sect.\ref{Experimental protocol}. 

\subsection{Sodium chloride (NaCl) solution}
\label{NaCl+H2O}
\begin{figure}
	\centering
	\includegraphics[width=\hsize]{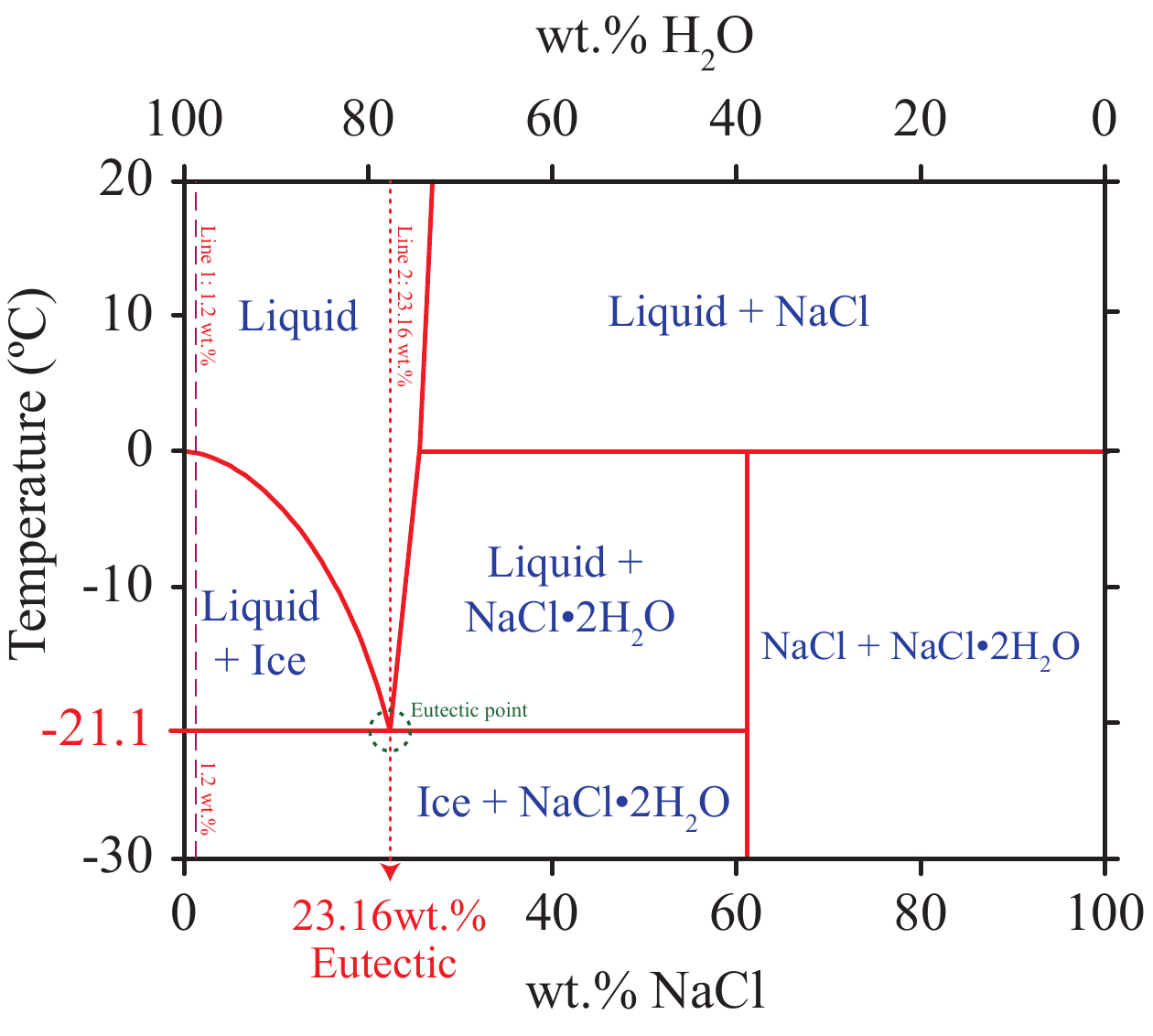}
	\caption[]{Phase diagram for the NaCl+H$_2$O solution. Redrawn after  \cite{PhasediagramNaClandMgCl2} and ref. therein.}
	\label{phasediagramNaCl+H2O}
\end{figure}

From the phase diagram provided in Fig.~\ref{phasediagramNaCl+H2O}, which is a conventional phase diagram of a NaCl solution, it is found which concentration of the solution will determine which solids will precipitate and the temperature at which equilibrium precipitation first occurs. A phase diagram displays the phases in a multi-component system as a function of the concentration (wt.\%) of the components and temperature. The selection of solution concentrations for the freezing/warming experiments was made based on the concentrations at which major phase changes occur in the phase diagram, such as the eutectic temperature and transitions between salt hydrates. The phase diagram can be examined following different vertical lines (dashed red) to illustrate the responses expected with different concentration in our experiments. 

From Fig.~\ref{phasediagramNaCl+H2O}, Line 1: 1.2 wt.\% describes a temperature change from 20$^{\circ}$C to -30$^{\circ}$C for dilute 1.2 wt.\% NaCl+H$_2$O solution; decreasing the temperature starting from 20$^{\circ}$C causes no change in the solution until the temperature goes below the red solid line at slightly below 0$^{\circ}$C, at this point pure H$_2$O ice starts to develop in the solution and the concentration of the liquid (unfrozen solution) increases until the temperature reaches the eutectic temperature of -21.1$^{\circ}$C for NaCl+H$_2$O. Further down this temperature, no more liquid will be found as it   will be pure water  ice and the eutectic solid mixture containing pure water ice (77wt.\%) and  NaCl$\cdot$2H$_2$O crystals known as hydrohalite (23wt.\%).
Another solution with a eutectic concentration of 23.16 wt.\%  of NaCl salt will follow Line 2 of Fig.~\ref{phasediagramNaCl+H2O}, precipitation of both ice and NaCl$\cdot$2H$_2$O occurs at the -21.1$^{\circ}$C eutectic temperature.
Therefore, two frozen samples with different concentrations of NaCl,  1.2 wt.\%  and eutectic (23.16 wt.\%), were manufactured and their temperatures, thermal conductivities, and calorimetry were measured. The results are compared with the pure water sample introduced in Sect.~ \ref{pure water}.
As mentioned in Sect.~\ref{Experimental protocol}, a small sample of the liquid salt-water solution was set aside for the calorimetry measurements.
\FloatBarrier

\begin{figure}
	\centering
	\includegraphics[width=\hsize]{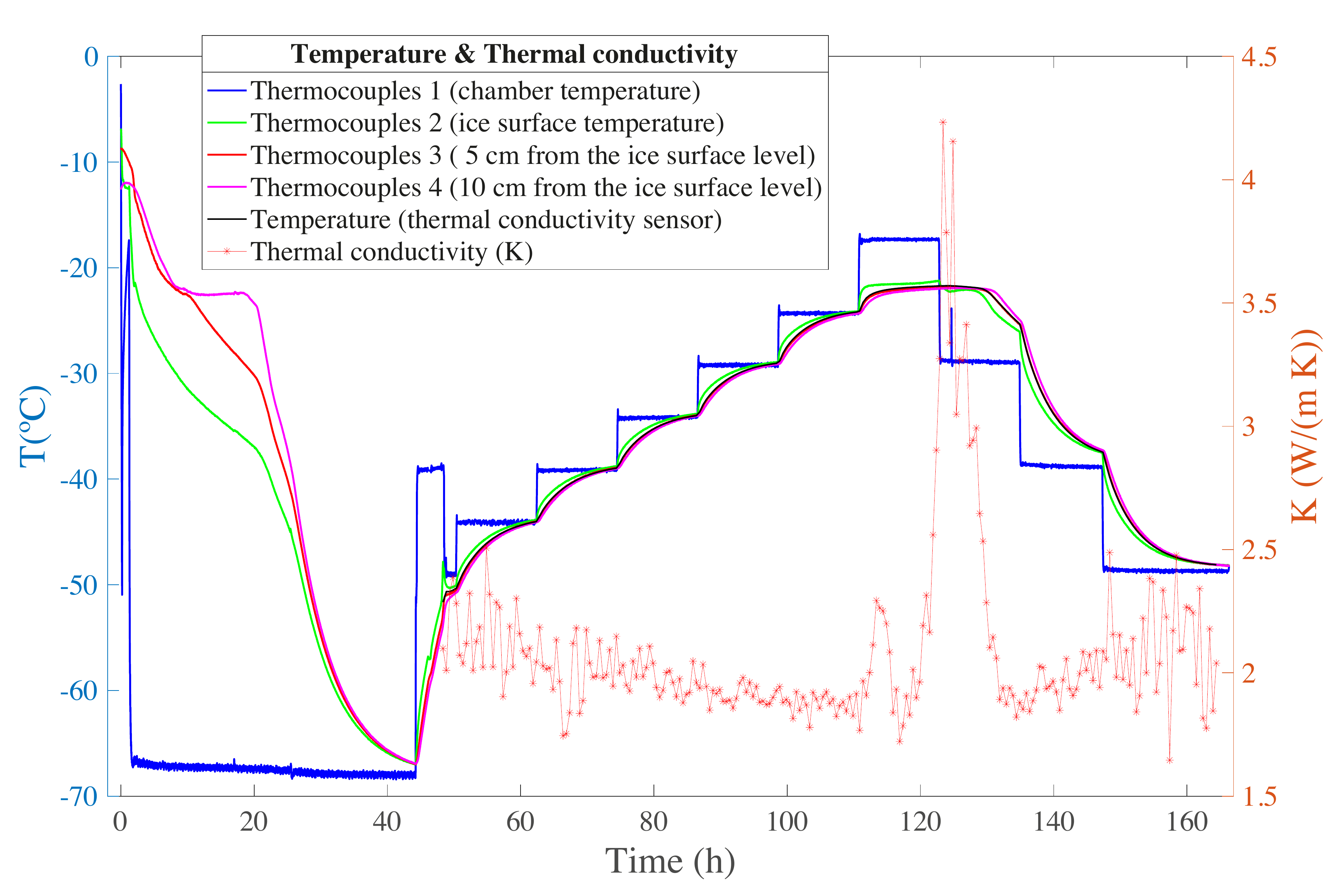}
	\caption{Thermal conductivity (right axis) and temperature (left axis) values measured by thermocouples and thermal conductivity meter for eutectic (23.16 wt.\%) NaCl + H$_2$O ice as a function of time. To link the temperature values to their corresponding thermal conductivity values, for instance at 100 h, an imaginary vertical line should be projected from the thermal conductivity value to the temperature values on top.}
	\label{TemperatureNaCl}
\end{figure}

\begin{figure}
	\centering
	\includegraphics[width=\hsize]{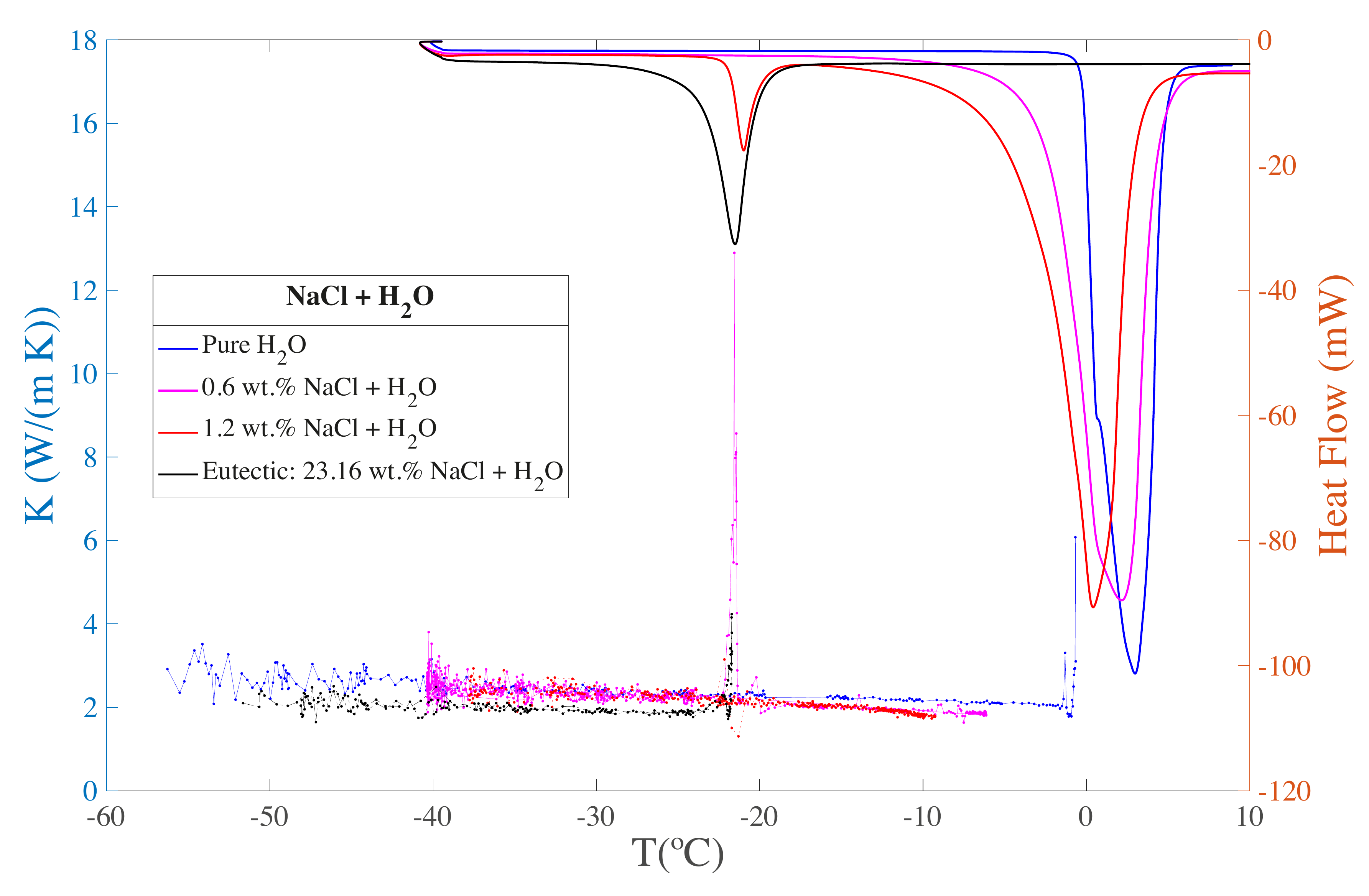}
	\caption{Thermal conductivity (left axis) and calorimetry (right axis) measurements for NaCl + H$_2$O with different NaCl concentration. Blue trace is pure H$_2$O. Red trace is 1.2 wt.\% NaCl + H$_2$O. Black trace is (Eutectic): 23.16 wt.\% NaCl + H$_2$O.}
	\label{NaCl_23_16wt}
\end{figure}

The onset temperature, $T_{onset}$, corresponds to the start of the peak in the heat flow (mW) measured by the calorimeter, see Fig.~\ref{NaCl_23_16wt}. This temperature coincides with the start of the crystallization during the freezing process of the solution. The onset temperature coincides with the plateau in Fig.~\ref{TemperatureNaCl} between 0 and 20 hours where the heat is being released as water ice continuously crystallises or freezes, which is the result of slowing down the cooling of the water solution.  
The crystallization would continue to take place until the liquid phase changes completely into solid phase. 
Furthermore, the temperature at which the first ice starts to melt is referred to as the onset temperature during the melting process. The melting of the ice solution would continue to take place until the solid changes completely into the liquid phase. This corresponds to a long period of constant temperature recorded by the thermocouples due to the endothermic melting in Fig.~\ref{TemperatureNaCl} between 155 and 160 hours.

From Figs.~\ref{phasediagramNaCl+H2O} and ~\ref{NaCl_23_16wt} the following information could be extracted: 
\begin{enumerate}
	\item Blue line in Fig.~\ref{NaCl_23_16wt} (Pure water):
		\begin{itemize}
			\item T$_{onset}$ = -0.21$^{\circ}$C --> 100\% water ice (pure H$_2$O).
		\end{itemize}
		
	\item Red line in Fig.~\ref{NaCl_23_16wt} (1.2 wt.\%  NaCl + H$_2$O):
			\begin{itemize}
				\item T$_{onset}$ = -21.8$^{\circ}$C -->  1.2\% eutectic mixture hydrohalite (23 wt.\% NaCl.H$_2$O)  + pure water ice (77 wt.\% water ice H$_2$O).
				\item T$_{onset}$ = -2.6$^{\circ}$C --> 98.8\% pure water ice. 
			\end{itemize}
			
	\item Black line in Fig.~\ref{NaCl_23_16wt} (23.16  wt.\%  NaCl + H$_2$O):
			\begin{itemize}
				\item T$_{onset}$ = -23.4$^{\circ}$C --> 100\%  eutectic mixture hydrohalite (23 wt.\% NaCl.H$_2$O)  + pure water ice (77 wt.\% water ice H$_2$O).
			\end{itemize}
\end{enumerate}

It can be noted that around -21$^{\circ}$C there is a peak in the thermal conductivity measurement with NaCl salt concentrations (1.2wt.\% and 23wt.\%), see Fig.~\ref{NaCl_23_16wt}. This peak is due to the fact that at this temperature there is a phase change in the NaCl + H$_2$O solutions. The heat release in the ice as a consequence of this phase change leads to a false value of the thermal conductivity. This peak is, nevertheless, a sensitive indicator of phase changes in the salt ice. This peak matches the phase diagram of the NaCl + H$_2$O system ( Fig.~\ref{phasediagramNaCl+H2O}). 

Fig.~\ref{NaCl_23_16wt} compares the thermal conductivities and the calorimetry of H$_2$O and  NaCl with different concentrations. It can be appreciated that the thermal conductivity of the eutectic ice mixture 23.16  wt.\% NaCl + H$_2$O  (black dots) is slightly different than the thermal conductivity of the other samples,  1.2wt.\% NaCl + H$_2$O (red dots) and pure water  (blue dots), which are very similar to each other. This figure illustrates what has already been pointed out about the calorimetric measurements and the phase diagram of NaCl + H$_2$O  system: both solutions behave similarly to pure water:
\begin{enumerate}
	\item Pure water (H$_2$O)  (blue dots) --> 100\% water ice + 0\% eutectic mixture (23wt.\% NaCl.H$_2$O  + 77 wt.\% H$_2$O ice).
	\item 1.2\% M NaCl + H$_2$O  (red dots) --> 98.8 \% water ice + 1.2 \% eutectic mixture (23wt.\% NaCl.H$_2$O  + 77 wt.\% H$_2$O ice).
\end{enumerate}

\subsection{Magnesium sulphate (MgSO$_4$) solution} \label{MgSO4+H2O}
\begin{figure}
	\centering
	\includegraphics[width=\hsize]{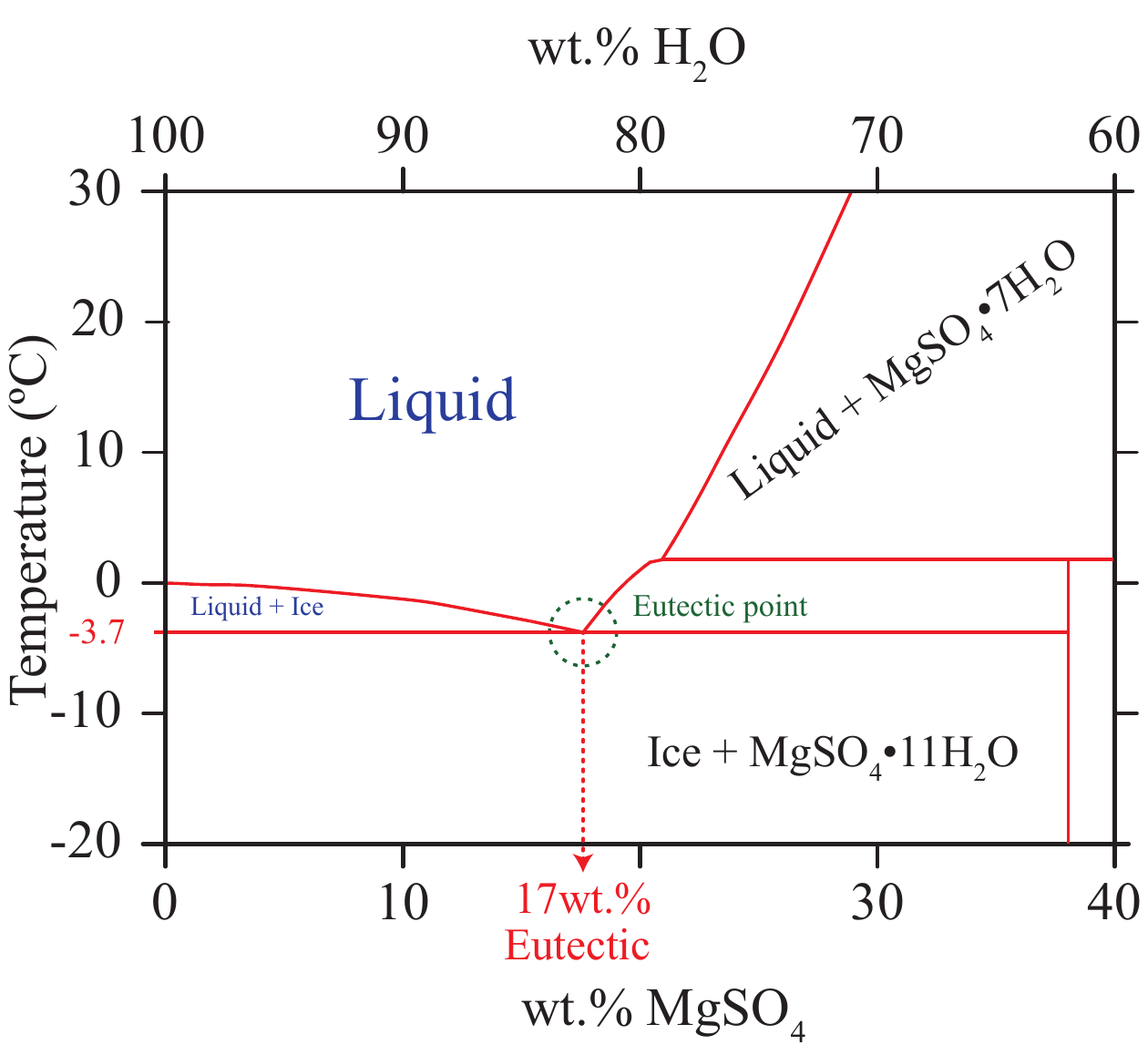}
	\caption[]{Phase diagram for the eutectic  17 wt.\% MgSO$_4$ + H$_2$O solution. Redrawn after \cite{PhasediagramMgSO4} and ref. therein.}
	\label{phasediagramMgSO4H2O}
\end{figure}

\begin{figure}
	\centering
	\includegraphics[width=\hsize]{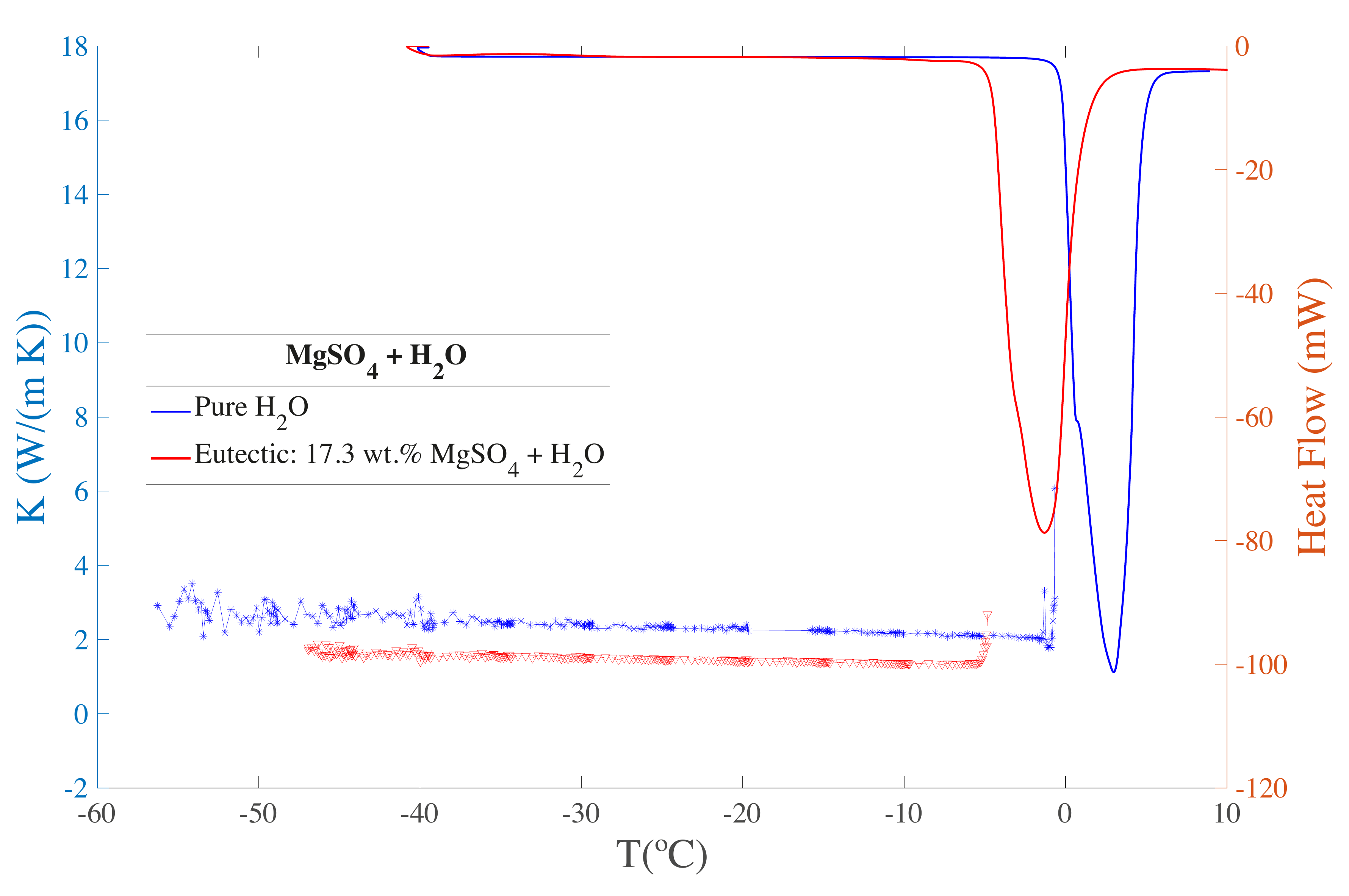}
	\caption{Thermal conductivity (left axis) and calorimetry (right axis) measurements for MgSO$_4$ + H$_2$O with eutectic MgSO$_4$ concentration. Blue trace is pure H$_2$O. Red trace is eutectic 17.3 wt.\% MgSO$_4$+ H$_2$O.}
	\label{MgSO4 17.3wt}
\end{figure}

From the phases diagram, it could be found that the concentration of the solution will determine which solids precipitate and the temperature at which equilibrium precipitations first occurs. 
From Fig.~\ref{phasediagramMgSO4H2O}, the equilibrium phase diagram of MgSO$_4$ shows that a dilute solution will precipitate to form ice near 0$^{\circ}$C, while a solution with a eutectic concentration of 17 wt.\%  salt  will precipitate both ice and MgSO$_4\cdot$11H$_2$O at the -3.7$^{\circ}$C eutectic temperature. Therefore, we performed one run using solutions at 17 wt.\% salt concentration, which is the eutectic composition of the binary system MgSO$_4\cdot $H$_2$O. The onset temperature obtained from the calorimetry measurement for this sample is T$_{onset}$ = - 4.48$^{\circ}$C , see Fig.~\ref{MgSO4 17.3wt}.

\FloatBarrier

\begin{figure}
	\centering
	\includegraphics[width=\hsize]{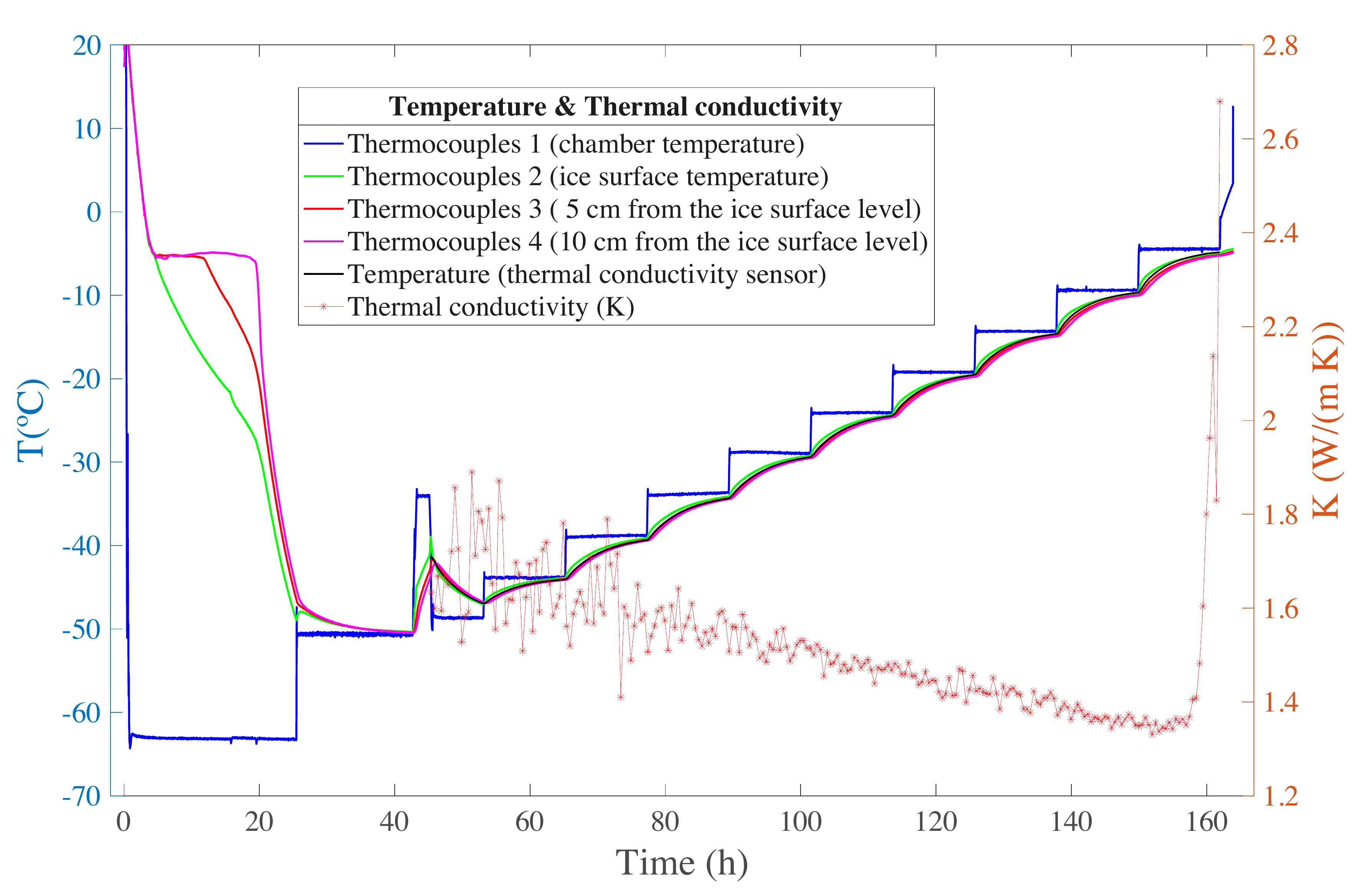}
	\caption{Thermal conductivity (right axis) and temperature (left axis) values measured by thermocouples and thermal conductivity meter for eutectic (17 wt.\%) MgSO$_4 + $H$_2$O ice as a function of time. To link the temperature values to their corresponding thermal conductivity values, for instance at 100 h, an imaginary vertical line should be projected from the thermal conductivity value to the temperature values on top.}
	\label{TemperatureMgSO4-1}
\end{figure}

Fig.~\ref{TemperatureMgSO4-1} shows the thermal conductivity and the temperature values measured by thermocouples and thermal conductivity meter for MgSO$_4 + $H$_2$O ice as a function of time. It can be noted that the steps in this figure are longer, taking more time to reach the temperature due to the lower thermal conductivity of this ice mixture. A peak at -3.7$^{\circ}$C in the thermal conductivity measurement (red traces) can be seen. This peak coincides with the phase change from liquid to solid in the MgSO$_4 + $H$_2$O ice mixture. 
A comparison of the thermal conductivity and calorimetry results between pure water and the magnesium sulfate sample is shown in Fig.~\ref{MgSO4 17.3wt}. The conductivity values are lower in the mixture with magnesium sulfate than with only pure water.
\FloatBarrier

\subsection{Sodium sulphate (Na$_2$SO$_4$) solution}
\label{Na2SO4+H2O}
 
\begin{figure}
	\centering
	\includegraphics[width=\hsize]{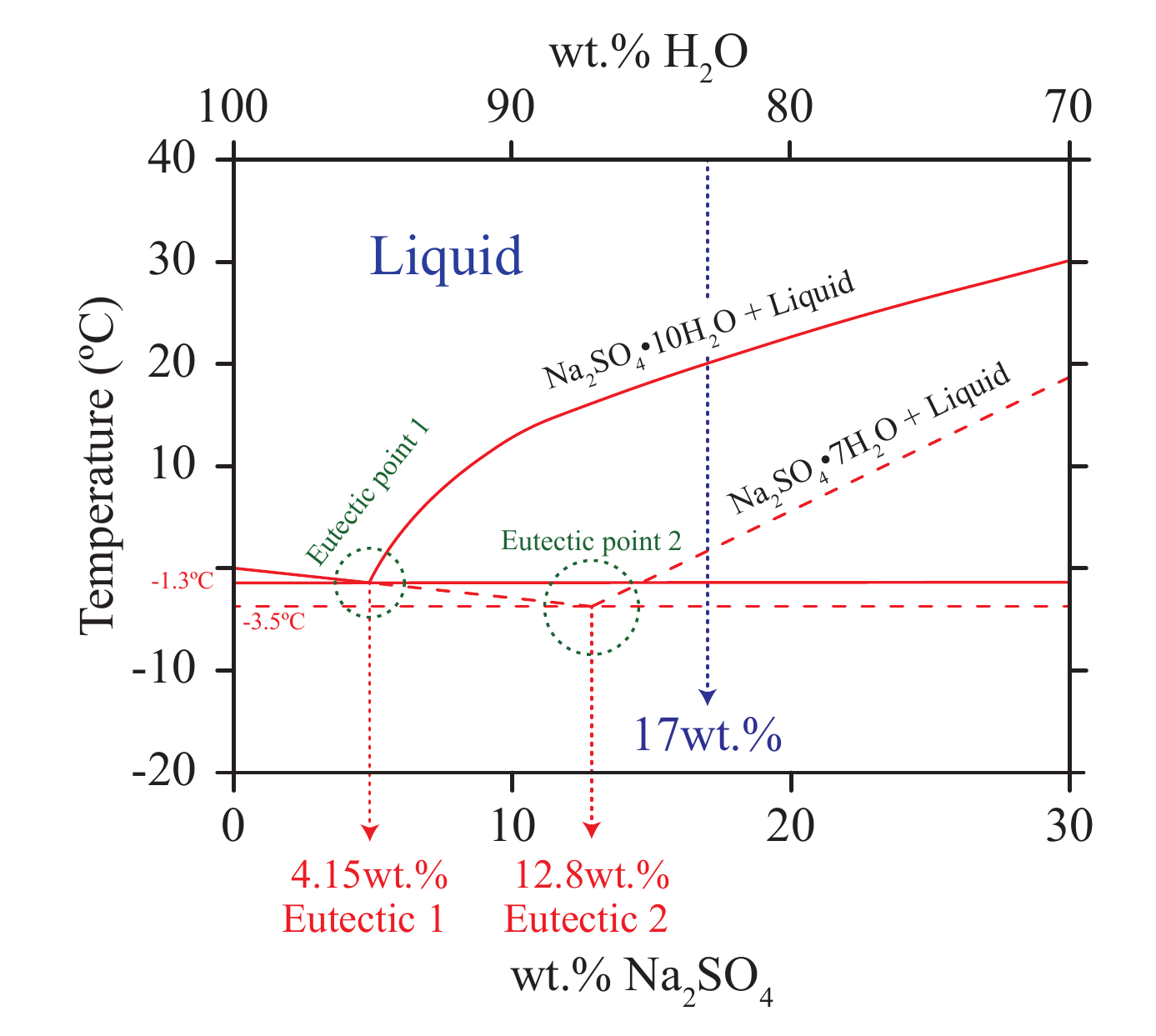}
	\caption{Phase diagram for the Na$_2$SO$_4$ + H$_2$O solution showing stable phase boundaries (solid) and metastable phase boundaries (dashed). E$_1$ and E$_2$ are the mirabilite–ice and the Na$_2$SO$_4\cdot7$ H$_2$O–ice eutectics, respectively. Redrawn after \citep{PhaseNA2SO4} and ref. therein.}
	\label{phasediagramNa2SO4H2O}
\end{figure}

Sodium sulphate decahydrate or mirabilite Na$_2$SO$_4\cdot$10 H$_2$O  is the stable phase in contact with an equilibrium mixture of  Na$_2$SO$_4$ and H$_2$O  at room temperature and atmospheric pressure. The eutectic, E$_1$ in Fig.~\ref{phasediagramNa2SO4H2O}, between mirabilite and ice I$_h$ (hexagonal ice crystal) is at -1.3$^{\circ}$C, 4.15 wt.\% Na$_2$SO$_4$ + H$_2$O. A metastable phase, Na$_2$SO$_4\cdot7$ H$_2$O , is known at room temperature and atmospheric pressure. The metastable heptahydrate ice I$_h$ eutectic, E$_2$ in Fig.~\ref{phasediagramNa2SO4H2O}, is at -3.55$^{\circ}$C, 12.8 wt.\% Na$_2$SO$_4$. 

Therefore, a set of three experiments has been performed using $E_1$ concentration, 4.15 wt.\%, $E_2$ concentrations  12.8 wt.\% and 17 wt.\%  of Na$_2$SO$_4$, see Fig.~\ref{phasediagramNa2SO4H2O}. The latter experiment (17 wt.\%  of Na$_2$SO$_4$) was performed to check the sensibility of the thermal conductivity measurement during the phase change and the concentration effect on the thermal conductivity.

\begin{figure}
	\centering
	\includegraphics[width=\hsize]{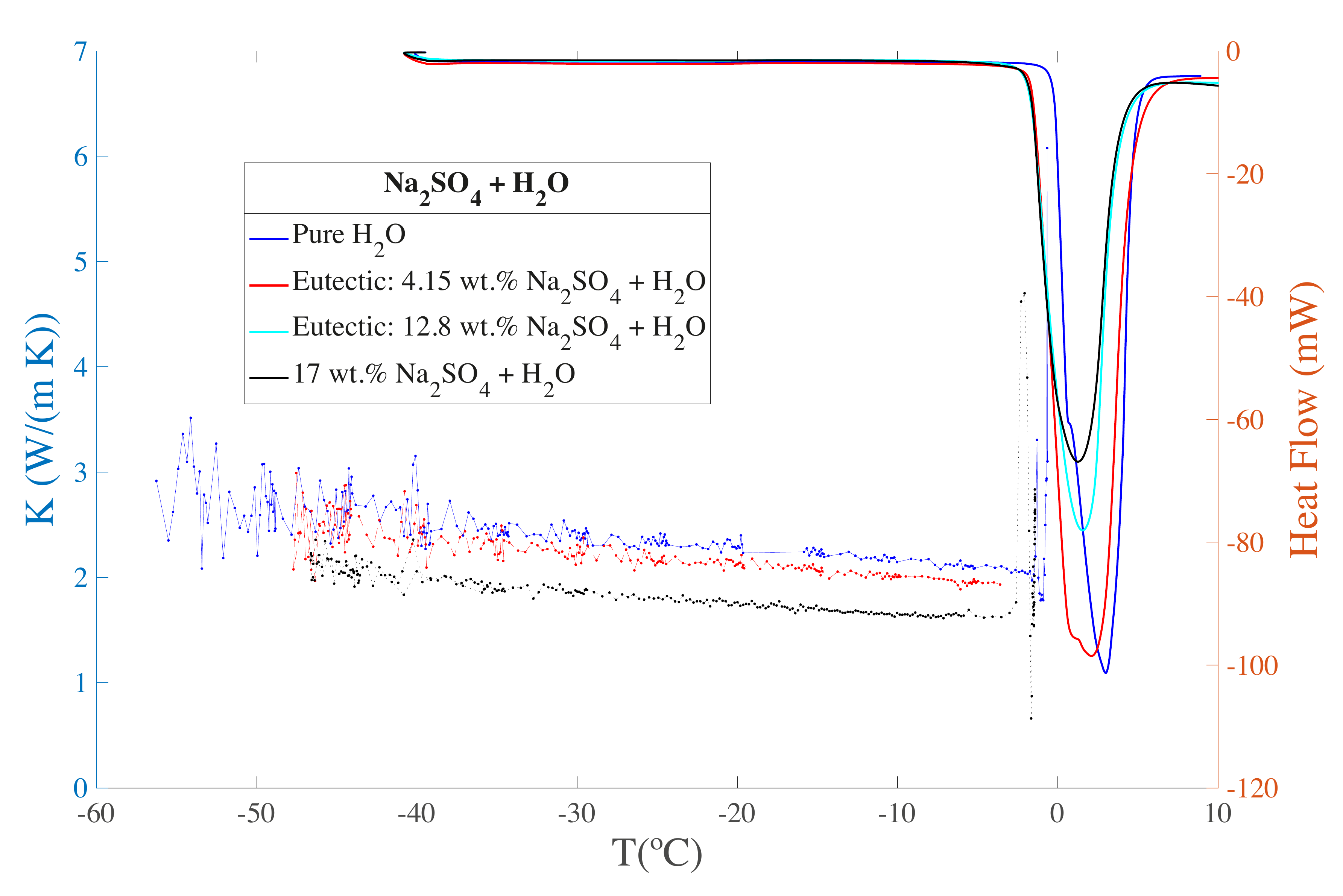}
	\caption{Thermal conductivity (left axis) and calorimetry (right axis) measurements for Na$_2$SO$_4$ + H$_2$O with different Na$_2$SO$_4$ concentration. Blue trace is pure H$_2$O. Red trace is 4.15 wt.\% Na$_2$SO$_4$ + H$_2$O. Black trace is 17 wt.\% Na$_2$SO$_4$ + H$_2$O.}
	\label{Na2SO4_17wt}
\end{figure}

The calorimetry measurements in Fig.~\ref{Na2SO4_17wt} show that the melting points are quite similar for the three concentrations, 4.15 wt.\%, 12.8 wt.\% and 17 wt.\%, their values fall around -1.5$^{\circ}$C. Therefore, it can be concluded that only stable phase of  sodium sulphate decahydrate or mirabilite Na$_2$SO$_4\cdot$10 H$_2$O  is formed with 10 water molecules and no evidence of the metastable phase, Na$_2$SO$_4\cdot$7 H$_2$O, which would be reflected in a melting point at a lower temperature, around -3.5$^{\circ}$C in the calorimetry measurements, as it can be seen in the phase diagram in Fig.~\ref{phasediagramNa2SO4H2O}. However, in the thermal conductivity measurements, Fig.~\ref{Na2SO4_17wt}, it could be noted that the peak of the 17 wt.\% Na$_2$SO$_4$ solution is lower that the peak of the 4.15 wt.\% Na$_2$SO$_4$.

\begin{itemize}
	\item T$_{onset 1}$ = -1.37$^{\circ}$C (4.15 wt.\%) 
	\item T$_{onset 2}$ = - 1.6$^{\circ}$C (12.8 wt.\%)
	\item T$_{onset 3}$ = - 1.62$^{\circ}$C (17 wt.\%)  
\end{itemize}
\FloatBarrier

\begin{figure}
	\centering
	\includegraphics[width=\hsize]{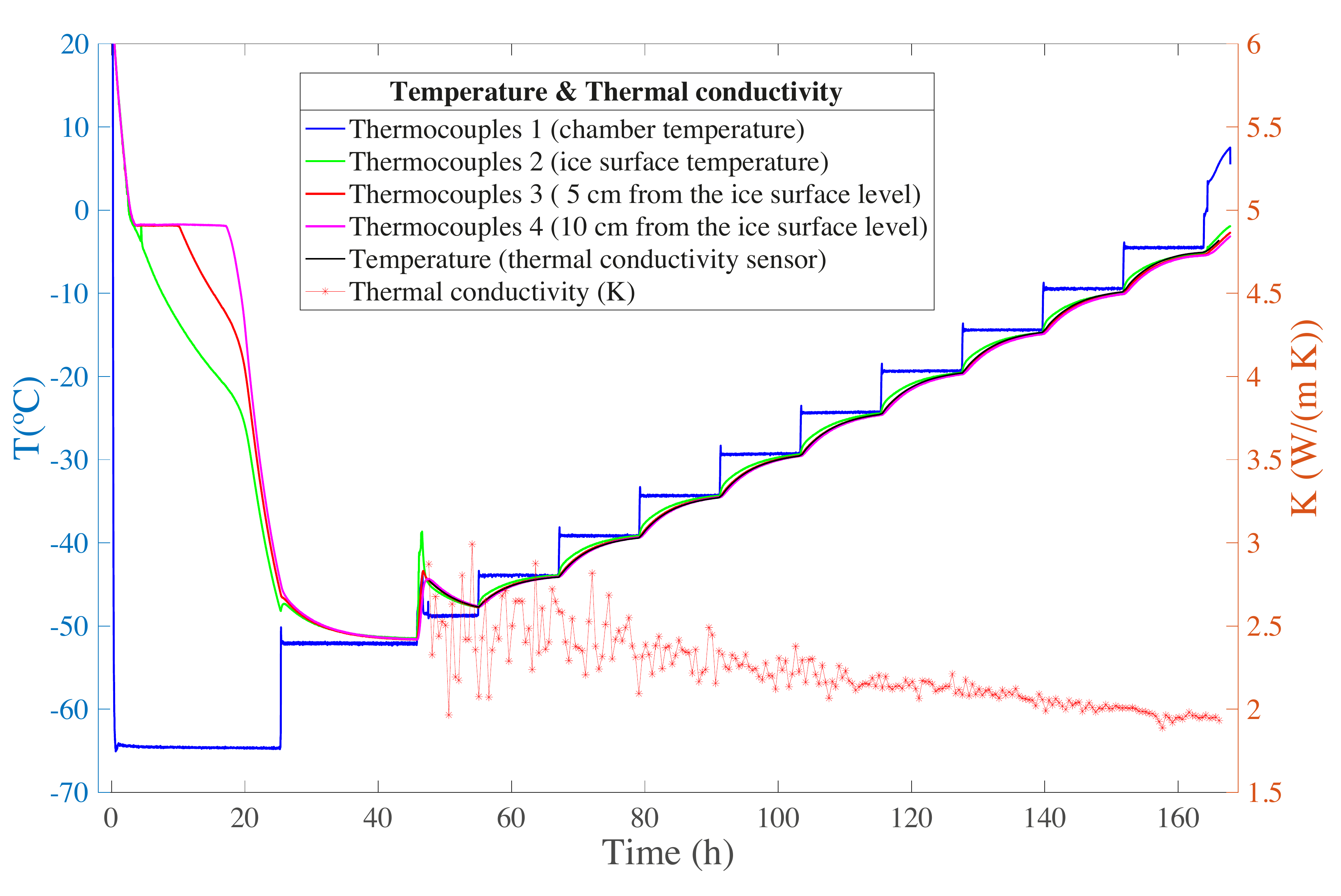}
	\caption{Thermal conductivity (right axis) and temperature (left axis) values measured by thermocouples and thermal conductivity meter for eutectic$_1$ (4.15 wt.\%) Na$_2$SO$_4$ + H$_2$O ice as a function of time. To link the temperature values to their corresponding thermal conductivity values, for instance at 100 h, an imaginary vertical line should be projected from the thermal conductivity value to the temperature values on top.}
	\label{TemperatureNa2SO4}
\end{figure}

\subsection{Magnesium chloride (MgCl$_2$) solution}
\label{MgCl+H2O}
\begin{figure}
	\centering
	\includegraphics[width=\hsize]{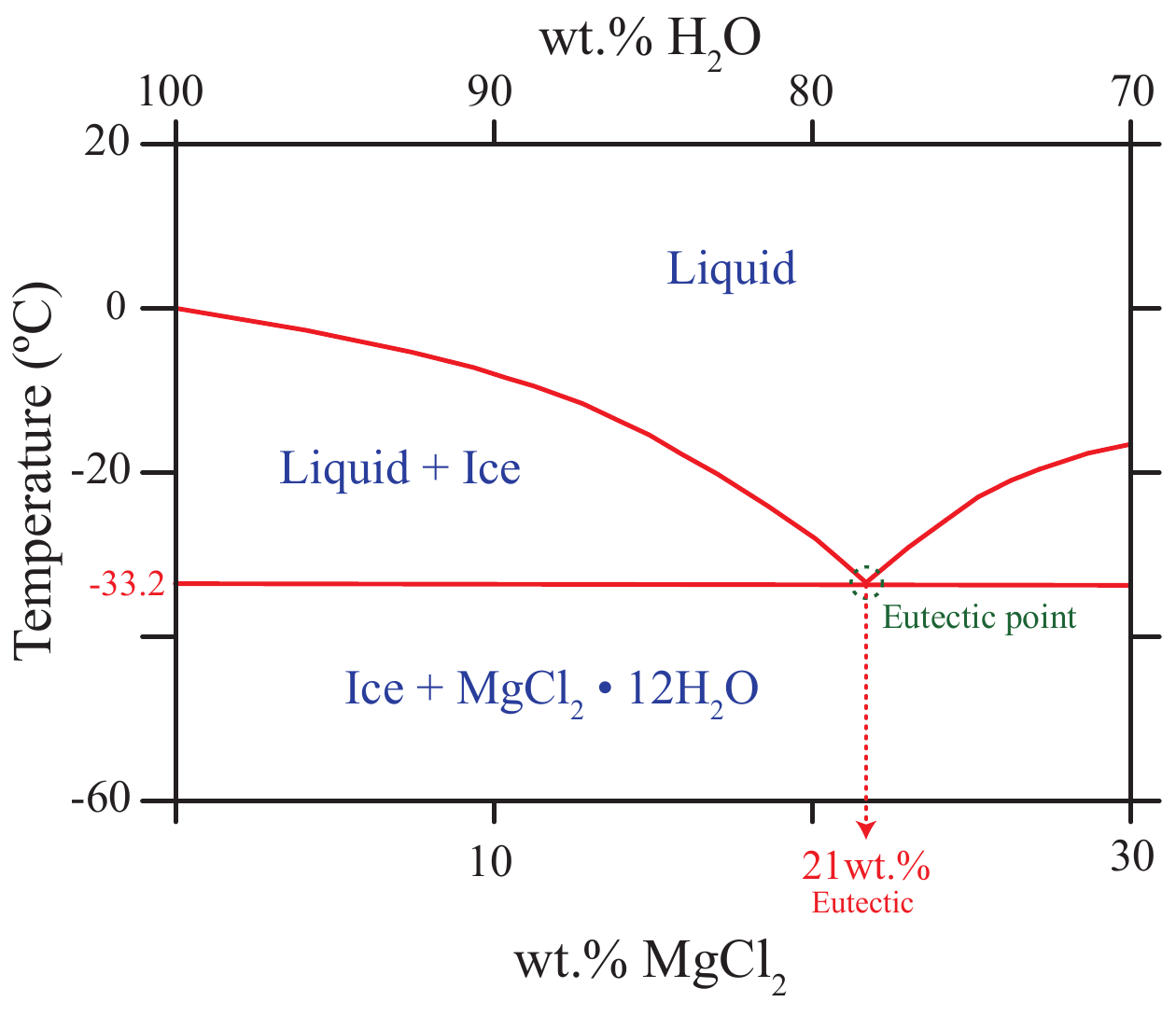}
	\caption[]{Phase diagram for the MgCl$_2$+H$_2$O solution. Redrawn after  \cite{PhasediagramNaClandMgCl2} and ref. therein.}
	\label{phasediagramMgCl2+H2O}
\end{figure}

\begin{figure}
	\centering
	\includegraphics[width=\hsize]{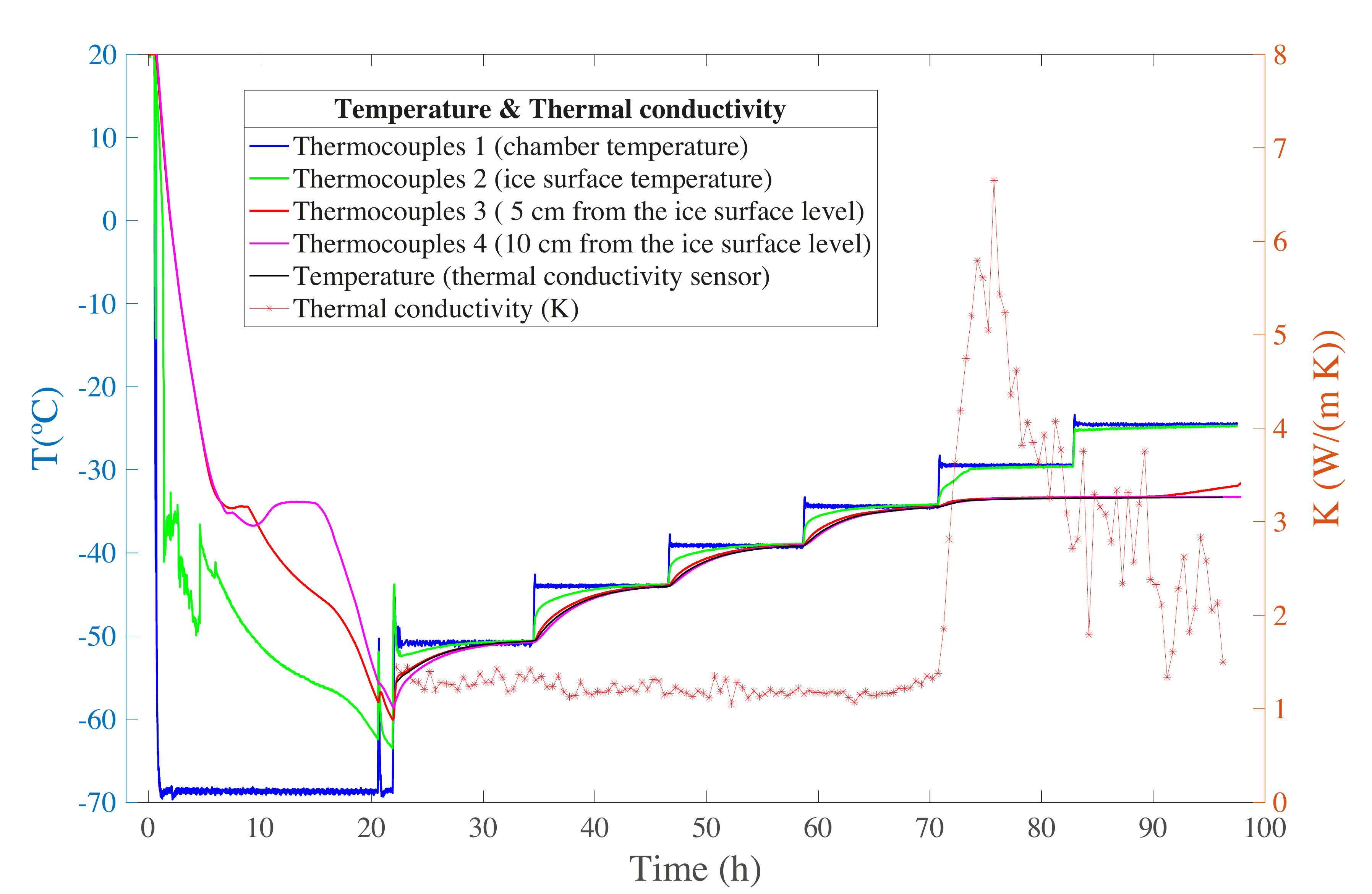}
	\caption{Thermal conductivity (right axis) and temperature (left axis) values measured by thermocouples and thermal conductivity meter for the eutectic (21 wt.\%) MgCl$_2 + $H$_2$O ice as a function of time. To link the temperature values to their corresponding thermal conductivity values, for instance at 100 h, an imaginary vertical line should be projected from the thermal conductivity value to the temperature values on top.}
	\label{TemperatureMgCl2}
\end{figure}

From Fig.~\ref{phasediagramMgCl2+H2O}, the equilibrium phase diagram of MgCl$_2$ shows that a dilute solution will precipitate ice near 0$^{\circ}$C, while a solution with a eutectic concentration of 21 wt.\%  of MgCl$_2$ salt concentration will precipitate both ice and MgCl$_2\cdot$12H$_2$O at the -33.2$^{\circ}$C eutectic temperature. Therefore the eutectic stable phase occurs around 21 wt.\% of MgCl$_2$ salt concentration. Fig.~\ref{TemperatureMgCl2} presents the temperature values measured by the thermocouples (left axis) and thermal conductivity (right axis) meter of the eutectic solution as a function of time. This figure shows a peak around -33$^{\circ}$C in the thermal conductivity measurement (red trace with dots). This peak coincides with the phase change from liquid to solid in MgCl$_2$ + H$_2$O (Fig.~\ref{phasediagramMgCl2+H2O}). 

\begin{figure}
	\centering
	\includegraphics[width=\hsize]{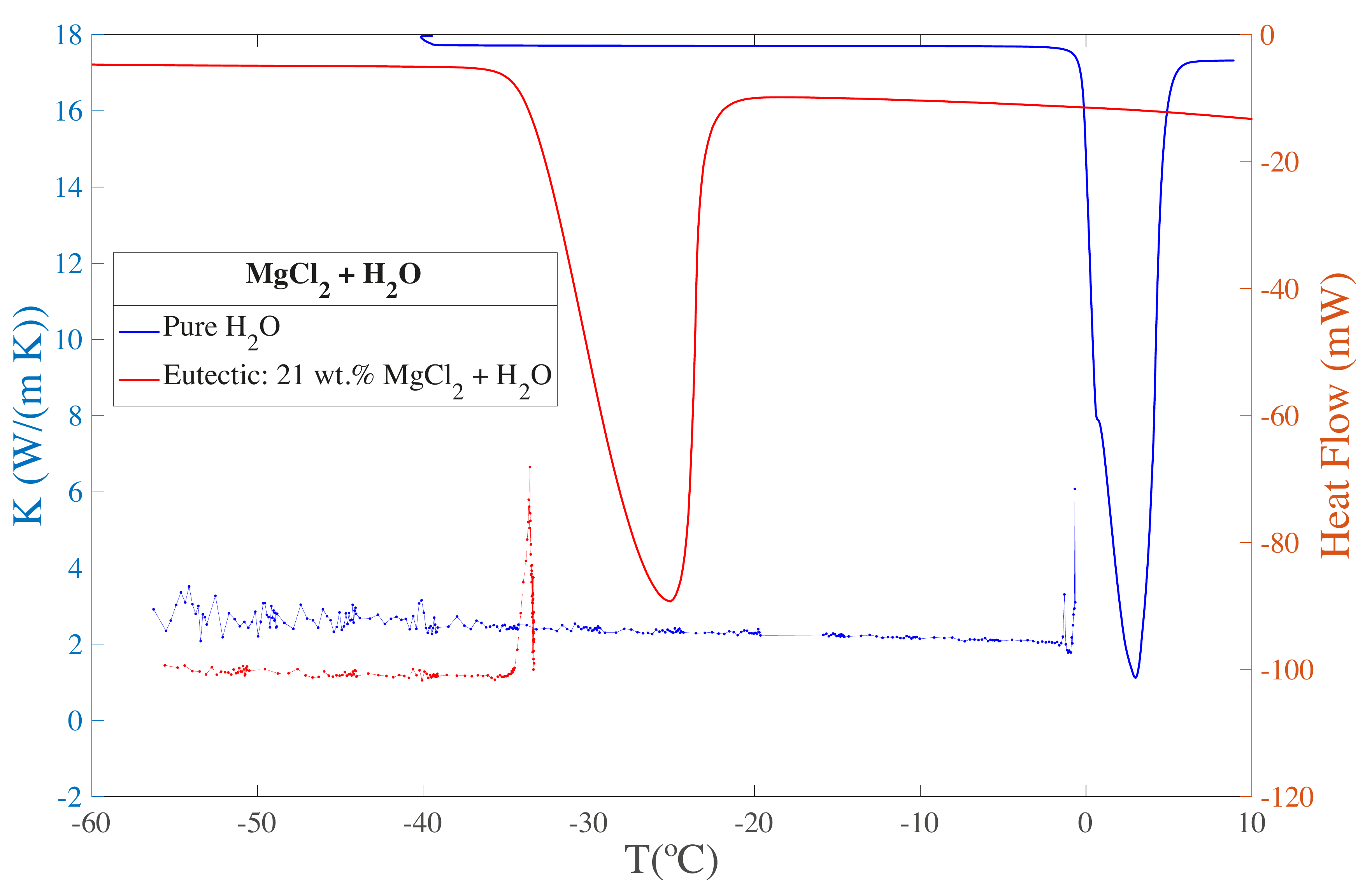}
	\caption{Calorimetry measurements for MgCl$_2$ + H$_2$O with the eutectic 21 wt.\% MgCl$_2$ concentration. Blue trace is pure H$_2$O. Red trace is eutectic 21 wt.\% MgCl$_2$ + H$_2$O.}
	\label{MgCl_2_17}
\end{figure}

The MgCl$_2$ + H$_2$O system can supercool \citep{TONER2014} below its eutectic temperature at -33.2$^{\circ}$C before crystallization occurs. That is, the temperature drops below the freezing point for some time, as shown by the little dip in the cooling curve, magenta trace, of the Fig.~\ref{TemperatureMgCl2} between 7 and 12 hours, below -33.2$^{\circ}$C. This region corresponds to an unstable form of the magnesium chloride (MgCl$_2$) solution, it does not crystallise well, it solidifies in a disorganized way, and it does so at temperatures well below freezing. If the magnesium chloride (MgCl$_2$) solution is allowed to stand, if cooling is continued, or if a small crystal of the solid phase is added (a seed crystal), the supercooled liquid solution will convert to a solid, sometimes quite suddenly. As the solution freezes, the temperature increases slightly due to the heat evolved during the freezing process and then holds constant at the melting point as the rest of the solution freezes. Therefore it needs to be super-cooled to crystallize well. The calorimeter used for the previous experiments can only cool down to -40$^{\circ}$C, which explains why in our previous attempts (not shown in this paper) only water ice crystallized. Hence a longer cycle and heating ramp at 0.5 K/min was selected, instead of 1 K/min as in the previous experiments. It appears that melting at -33$^{\circ}$C is finally observed, and water ice at 0$^{\circ}$C is also evident. The eutectic at -33$^{\circ}$C is not very well defined, this may be due to the fact that the salt solution did not crystallize well and partially remained as an amorphous phase, or because the reaction is highly exothermic and therefore gives off a lot of heat.
As previously mentioned, the thermal transitions were analyzed by Differential Scanning Calorimetry (DSC) on a Mettler Toledo DSC 822e calorimeter (Schwerzenbach) equipped with a liquid nitrogen accessory, see Fig.~\ref{MgCl_2_17}. In this case, the sample was cooled from 25$^{\circ}$C to -90$^{\circ}$C at a rate of 10$^{\circ}$C/min, maintained for 7 min at this temperature and reheated from -90$^{\circ}$C to 25$^{\circ}$C at a rate of 10$^{\circ}$C/min. The calorimetry measurements for the eutectic 21 wt.\% MgCl$_2$ + H$_2$O (red trace) and pure water (blue trace) are shown in Fig.~\ref{MgCl_2_17}.

\FloatBarrier

\section{Discussion}
\label{Discussion}
\begin{figure}
	\centering
	\includegraphics[width=\hsize]{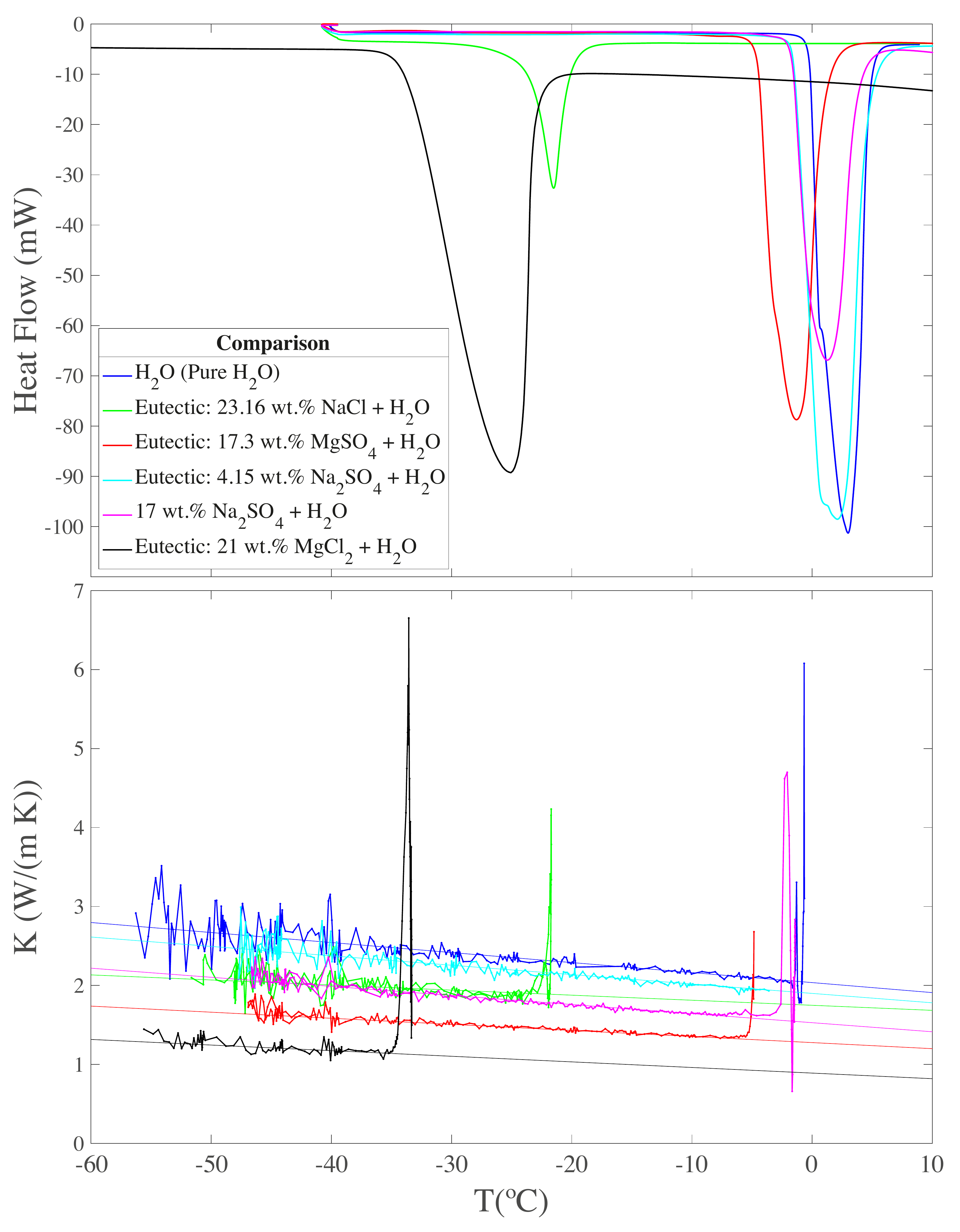}
	\caption{Thermal conductivity and calorimetry measurements for  H$_2$O (blue trace), eutectic 23.16 wt.\% NaCl + H$_2$O (green trace), eutectic 17.3 wt.\% MgSO$_4$ + H$_2$O (red trace), eutectic $4.15$ wt.\% Na$_2$SO$_4$ + H$_2$O (cyan trace), 17 wt.\% Na$_2$SO$_4$ + H$_2$O (magenta trace) and eutectic 21 wt.\% MgCl$_2$+ H$_2$O (black trace). Fit type: linear: a*T + b}. 
	\label{comparisonTC_CAL}
\end{figure}

Fig.~\ref{comparisonTC_CAL} shows a summary of the thermal conductivity (bottom panel) and calorimetry (top panel) measurements for H2O (blue trace), eutectic 23.16 wt.\% NaCl + H$_2$O (green trace), eutectic 17.3 wt.\% MgSO$_4$ + H$_2$O (red trace), eutectic 4.15wt.\% Na$_2$SO$_4$ + H$_2$O (cyan trace), 17 wt.\% Na$_2$SO$_4$ + H$_2$O (magenta trace) and eutectic 21 wt.\% MgCl$_2$ + H$_2$O (black trace). 

A conclusion of this work is that the sensitivity of our conductivity measurements served to identify phase changes, as a sudden rise of the conductivity forming a sharp peak that reaches values above the range of the thermal conductivity sensor, 4 W/(m K). The peak value is not the real value of the conductivity, but it serves to spot phase changes. These peaks are caused by a change in the temperature during the phase change that the sensor interprets as a large variation of the thermal conductivity. In addition, for solutions at the phase change temperature, convection induces temperature variations and the thermal conductivity measurements may have large uncertainties.
Sometimes this method was more sensitive than the calorimeter itself to identify phase changes. As an example, we refer to Fig.~\ref{NaCl_23_16wt} where a peak in the thermal conductivity is measured, even with 0.6wt.\% concentration of NaCl, a very low value, while the calorimeter did not detect a phase change.

Fig.~\ref{comparisonTC_CAL} illustrates that the thermal conductivity of the macroscopic ice samples grown for this study are different from each other. However, all thermal conductivities share similarities, such as the slope values of the thermal conductivity (Fig.~\ref{comparisonTC_CAL} and Table~\ref{Resumen}) as a function of the temperature which varies from 1 to 3 W/(m$\cdot$K). In all cases we can see a similar linear trend of thermal conductivity as a function of the temperature. A very interesting fact is that the salts lower the thermal conductivity, which means that the thermal gradients will be weaker if there are salts and that the heat is lost more slowly.
It can also be seen that certain solutions are more similar to each other, such as pure water and 4.15 wt.\% sodium sulphate (4.15 wt.\% Na$_2$SO$_4$ + H$_2$O) with a$_{H_2O}$ = -0.01268 and a$_{Na_2SO_4}$ = -0.0119, or 23.16 wt.\% Na-chloride (23.16 wt.\% NaCl) and 17.3 wt.\% sodium sulphate (17.3 wt.\% Na$_2$SO$_4$ + H$_2$O) with slightly different slope values of -0.00641 and -0.0077, respectively. 
In the temperature range studied, -55 to 0$^{\circ}$C, the range of the thermal conductivity goes from 1 W/(m$\cdot$K) for the eutectic 21 wt.\% magnesium chloride (21 wt.\% MgCl$_2$ + H$_2$O) at around -33$^{\circ}$C to 3 W/(m$\cdot$K) for the pure water (H$_2$O) at around -55$^{\circ}$C. Therefore, the highest values of thermal conductivity correspond to pure water (H$_2$O) (blue trace in Fig.~\ref{comparisonTC_CAL}) and the lowest values to the 21 wt.\% magnesium chloride (21 wt.\% MgCl$_2$ + H$_2$O), the black trace in Fig.~\ref{comparisonTC_CAL}. 

The results for the various phases, states and the temperature range studied are in most cases described well by the linear equation used in this paper: $k = a \cdot T + b$. In the literature, numerous fitting adjustments for the experimental thermal conductivity data have been found, such as the one used by \cite{Andersson_2005}, $k = a/T+b+c\cdot T$, \cite{Slack_1980}, $k = a/T$,  or linear fitting \cite[and ref. therein.] {Bonales2017}. However, in all these cited cases, the thermal conductivity data are for pure H$_2$O ice \citep{Andersson_2018,WOLFENBARGER2021}. 

Our data are in agreement with slurry ice (\cite{melinder2003}) where the higher conductivity of ice compared to the liquid phase enhances the thermal conductivity of the system. Indeed, this author reports the thermal conductivity of ice slurries is clearly higher than that of single-phase aqueous solutions.

\begin{table}
	\centering
	\includegraphics[width=\hsize]{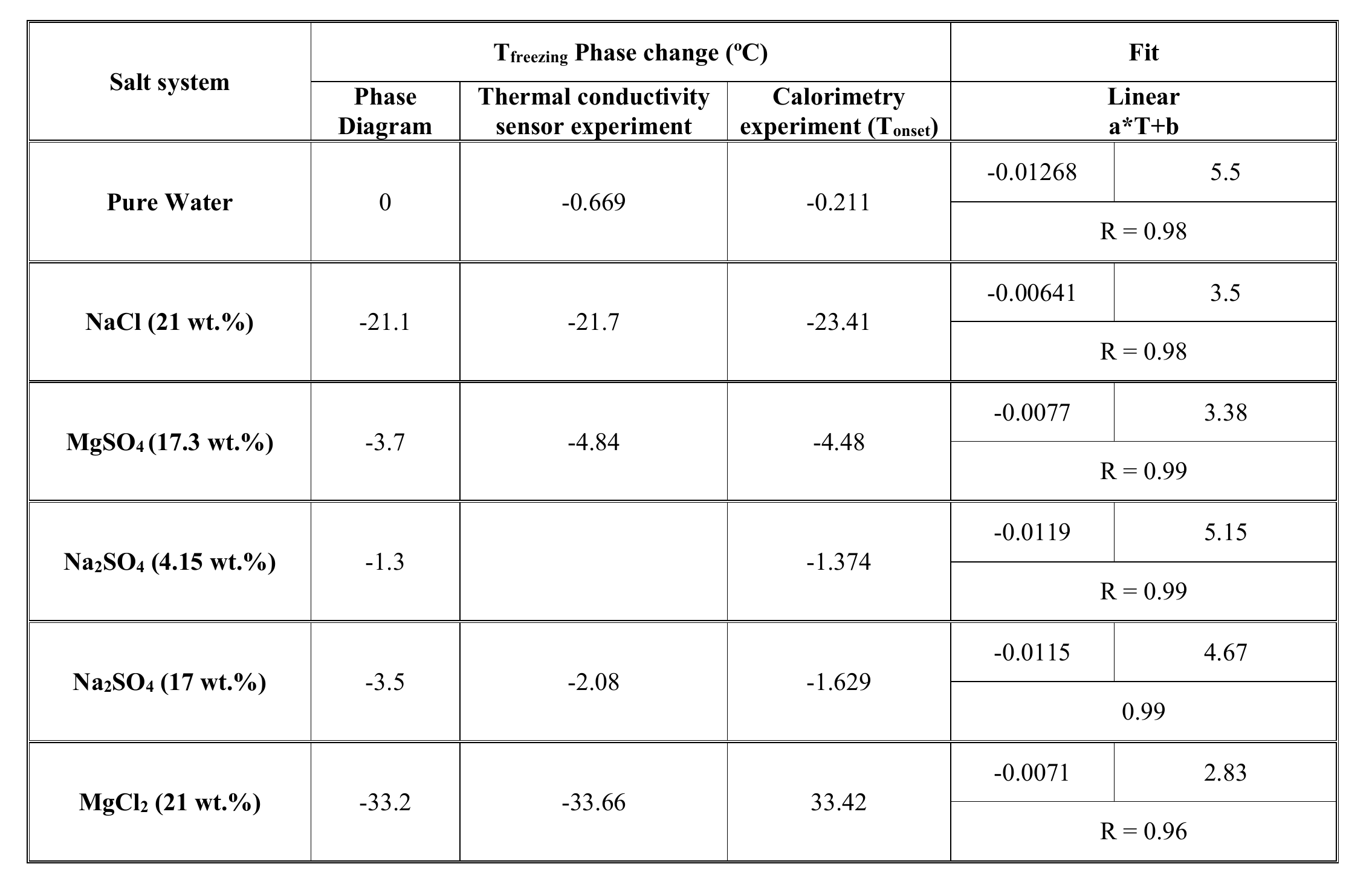}
	\caption{Summary of the literature (phase diagrams) and experimental (thermal conductivity sensor and calorimetry experiments) phase change freezing temperatures and linear fitting values for the different salt systems reported in this paper.} 
	\label{Resumen}
\end{table}

\section{Astrophysical implications and conclusions}
\label{astro_implications}
In this paper, we provide measurements of temperature, calorimetry and thermal conductivity in ice analogs containing salts with different concentrations as input for the thermodynamical models of Jovian icy moons that will be developed to interpret the volume of data from these missions. These measurements of thermal conductivity and calorimetry for relevant salt solutions are expected to be useful to better constrain the chemical composition, physical state, and temperature of the upper layers of Ganymede, Europa, and Callisto. Indeed, there is a significant lack of literature regarding the thermal properties of frozen salt solutions. 

The surfaces of these icy moons are exposed to cometary and meteoritic impacts. In particular, embedded micro-meteoroids may be relatively abundant in the icy crust and contribute to the thermal conductivity of the moon surfaces. However, the thermal conductivity values of meteorites cover a wide range, from 0.5 W/(m·K) in carbonaceous chondrites to more than 10 W/(m·K) in iron meteorites, and these values can strongly depend on temperature \citep{OPEIL_2010}. 

The MAJIS (Moons and Jupiter Imaging Spectrometer) instrument on board JUICE will perform visible-IR spectroscopy in the 0.4-5.7 $\mu$m range, with spectral resolution of 3-7 nm to characterize the composition and physical properties of these satellite's surfaces. The spatial resolution will be up to 25 m on Ganymede and about 100 km on Jupiter. The spectral resolution of MAJIS will be about 4 times better than the Galileo/Near-Infrared Mapping Spectrometer (NIMS) in the near infrared range (better than 7 nm/band between 2.25 and 5.54 $\mu$m), providing a data quality closer to lab spectra of several non-ice materials measured at temperatures representative of the icy Galilean satellites \cite{Piccioni_2019}. The ices, salts, minerals and organic materials will be observed in this spectral range. This mapping of the surface composition will serve to better understand the geological history of these moons. MAJIS has the capability to analyze the composition of non-water ice components and determine the state and crystallinity of water ice; for more information, see JUICE Red Book available online at \href{https://sci.esa.int}{https://sci.esa.int}.

Mapping of the moon temperatures will be provided by the Sub-millimeter Wave Instrument (SWI), along with the measurement of thermophysical and electrical properties of their surface/subsurface. The RIME (Radar for Icy Moons Exploration) instrument employs radar sounding to study subsurface structures of icy shells with 50-140 m resolution. GIS (Geographic Information System) system will combine spatially referenced information such as imaging, composition, sub-surface, topography, thermal and geophysical data to get a full picture of each moon.  
Using our dataset, which could be extended to lower temperatures in the near future with the design of a new thermal conductivity meter, the heat transfer and temperature in stratified moon (sub)surfaces can be simulated provided that the thickness, composition and structure of the different salt ice layers are characterized during the JUICE mission.     
\section{Data availability}
\label{Data availability}
The data underlying this article cannot be shared publicly. 

\section*{Acknowledgements}
\label{acknowledgements}
This project received financial support of The European Space Agency (ESA) contracts No.: RFP/3-15589/18/ES/CM and 4000126441/19/ES/CM: "Measurements of thermal and dielectric properties of ices in support to future radar measurements of Jovian Icy moons", The Unidad de Excelencia "Mar\'{i}a de Maeztu" MDM-2017-0737-- Centro de Astrobiolog\'{i}a (INTA-CSIC), The Spanish Ministry of Science, Innovation and Universities AYA2017-85322-R and PID2020-118974GB-C21 (AEI/FEDER, UE), Retos Investigaci\'{o}n BIA2016-77992-R (AEI/FEDER, UE), and "Explora Ciencia y Explora Tecnolog\'{i}a" [AYA2017-91062-EXP]. We are grateful to Anezina Solomonidou for assistance in the project proposal. The view expressed in this article can in no way be taken to reflect the official opinion of the European Space Agency. We thank the reviewer of this article for his constructive comments. 
\bibliographystyle{mnras}
\bibliography{GonzalezDiaz_mnras_TC} 
\label{biblio}
\end{document}